\documentclass[12pt]{article}
 \usepackage{epsfig}
 \usepackage{pstricks}

\def\d{\delta}
\def\e{\epsilon}

\def\l{\lambda}
\def\m{\mu}
\def\n{\nu}

\def\p{\pi}

\def\s{\sigma}
\def\t{\tau}

\def\cl{{\cal L}}

\def\bo{{\raise.15ex\hbox{\large$\Box$}}}               % D'Alembertian
\def\pr{\prod}                                          % product
\def\face{{\raise.2ex\hbox{$\displaystyle \bigodot$}\mskip-2.2mu \llap {$\ddot
        \smile$}}}                                      % happy face
\def\leftrightarrowfill{$\mathsurround=0pt \mathord\leftarrow \mkern-6mu
        \cleaders\hbox{$\mkern-2mu \mathord- \mkern-2mu$}\hfill
        \mkern-6mu \mathord\rightarrow$}       % <--> double differential
\def\dvec#1{\vbox{\ialign{##\crcr
        \leftrightarrowfill\crcr\noalign{\kern-1pt\nointerlineskip}
        $\hfil\displaystyle{#1}\hfil$\crcr}}}           % <--> accent
\def\beq{\begin{equation}}
\def\eeq{\end{equation}}

\def\beqx{\begin{displaymath}}
\def\eeqx{\end{displaymath}}

\def\beqa{\begin{eqnarray}}
\def\eeqa{\end{eqnarray}}
\def\NO{\nonumber}

\def\pl#1#2#3{Phys.~Lett.~{\bf B {#1}} (19{#2}) #3}
\def\np#1#2#3{Nucl.~Phys.~{\bf B {#1}} (19{#2}) #3}

\def\pr#1#2#3{Phys.~Rev.~{\bf D {#1}} (19{#2}) #3}

\catcode`@=11
\def\@citex[#1]#2{\if@filesw\immediate\write\@auxout{\string\citation{#2}}\fi
  \def\@citea{}\@cite{\@for\@citeb:=#2\do
    {\@citea\def\@citea{,\penalty\@m}\@ifundefined
      {b@\@citeb}{{\bf ?}\@warning
       {Citation `\@citeb' on page \thepage \space undefined}}%
\hbox{\csname b@\@citeb\endcsname}}}{#1}}

\def\citer{\@ifnextchar [{\@tempswatrue\@citexr}{\@tempswafalse\@citexr[]}}
\def\@citexr[#1]#2{\if@filesw\immediate\write\@auxout{\string\citation{#2}}\fi
  \def\@citea{}\@cite{\@for\@citeb:=#2\do
    {\@citea\def\@citea{--\penalty\@m}\@ifundefined
       {b@\@citeb}{{\bf ?}\@warning
       {Citation `\@citeb' on page \thepage \space undefined}}%
\hbox{\csname b@\@citeb\endcsname}}}{#1}}
\catcode`@=12
\begin{document}
\date{\mbox{ }}

\title{ 
{\normalsize     
\hfill \parbox{32mm}{DESY 98-188}}\\[25mm]
The Gauge-Higgs System\\
in Three Dimensions\\
to Two-loop Order\\[8mm]}
\author{
Frank Eberlein\thanks{e-mail: {\tt frankeb}@{\tt x4u2.desy.de}}\\
Deutsches Elektronen-Synchrotron DESY,      \\
Notkestr.\,85, D-22603 Hamburg, Germany}
\maketitle

\thispagestyle{empty}

\begin{abstract}
  \noindent The 3-dimensional gauge-Higgs system  
describes the non-perturbative infrared effects of the 
high-temperature phase
of the Standard Model.
We calculate the two-loop self-energies in the 3-dimensional $SU(2)$ Higgs model and in the corresponding  gauged non-linear 
$\sigma$-model.
As an application of the results, we estimate the dynamically generated vector boson mass
in the symmetric phase of the Higgs model by means of gap equations.

\end{abstract}

\newpage  

%\clearpage
%\pagenumbering{arabic}
%\setcounter{page}{1}
%\noindent
%\begin{minipage}[t]{16.5cm}
%\noindent
\section{Introduction}

The 3-dimensional gauge-Higgs system is interesting in many respects.
The phase structure of the Abelian model describes 
type-I and type-II
superconductors \cite{halp}, whereas the
 non-abelian $SU(2)$ Higgs model serves as 
the high-temperature limit of the corresponding 4-dimensional theory
and the electroweak Standard Model.
It is important for the study of the electroweak phase
transition 
and, since it contains all the infrared physics of the full theory,
for investigating non-perturbative effects, which are expected
in the high-temperature (symmetric) phase \cite{7-9o,vier}.
Already Feynman stressed the importance of examining
the 3-dimensional theory as a first step to understand confinement in
QCD \cite{fein}.
Both the 3- and 4-dimensional theories confine with a linear potential.

Lattice simulations of the Higgs model in 3 dimensions have shown
that a mass gap exists in the symmetric phase 
with a particle spectrum consisting of bound states of gluons with scalars
\cite{funf} and of bound states of 
gluons only (glueballs) \cite{ptw}.
In order to illuminate the connection between Higgs phase and 
symmetric phase in 3 dimensions, 
it is important to understand the behaviour of the
vector boson propagator.
A propagator mass for the vector boson
was studied on the lattice in \cite{kar}.
In this paper, we concentrate on an analytical treatment of the model, in
particular on propagator effects. 
We calculate the self-energy of the Higgs and vector boson
to two-loop order.
The one-loop self-energy for the vector boson
turns out to be
 dominated by the diagrams obtained in the corresponding non-linear 
$\s$-model \cite{BP}.
This is why we consider this simpler model first in our two-loop calculation.

%Its  inverse, the magnetic screening length, determines
%the size of non-perturbative effects in the symmetric phase
%and is closely related to the confinement scale of the effective 3-dimensional% theory which describes the high-temperature
%limit of the 4-dimensional finite temperature field theory \cite{funf}.
%In an apparently massless 3-dimensional
%Yang-Mills theory 
% the gauge coupling $g^2$ carries the dimension of mass, thereby
%providing a natural mass scale.

In the symmetric phase of the 3-dimensional 
$ SU(2) $ Higgs model, gap equations provide an analytical tool to calculate the vector boson propagator mass \cite{BP}. The gap equation is  a self-consistent equation
for the self-energy after resumming perturbation theory.
We have recently extended this method to two-loop order in the non-linear
$\sigma$-model \cite{frenk}.
Here, we solve the two-loop 
gap equation for the vector boson in the linear Higgs model
and compare the result with the two-loop gap mass in the non-linear case.
With the help of the two-loop
self-energies evaluated here, we show that the  
non-linear $\sigma$-model constitutes a reasonable approximation for infrared effects of the
linear Higgs model.

\bigskip
% Neglecting fermions, which do not incorporate new infrared
%physics, we restrict ourselves to an $SU(2)$ Higgs model and a non-linear
%$\s$-model.

In section \ref{c2}, 
the two-loop calculation of the vector boson self-energy in the non-linear $\s$-model is presented
 in unitary and in Feynman gauge.
The gauge-independence of the pole of the propagator is verified.
In section \ref{c5}, the corresponding calculations are done for the
Higgs and the vector field in the $SU(2)$ Higgs model.

 In section \ref{c6}, 
the two-loop gap equation is calculated in the $SU(2)$ Higgs model.
The analysis of the two-loop gap equation for the vector boson mass
in Feynman gauge 
suggests that the gap equation
approach is a reliable method to calculate the transverse
 propagator mass 
of the vector boson in the symmetric phase.

Appendix A summarizes the basic two-loop integrals.
In appendix B, the two-loop results of the non-linear $\s$-model 
are presented in more detail keeping the dimension arbitrary, and finally 
in appendix C, the two-loop 
Higgs and vector boson self-energy in the $SU(2)$ Higgs
model are  given 
in unitary and in Feynman gauge.
%and finally in appendix D, the two-loop  master 
%integral is evaluated numerically for several special mass cases.

%\input{ip}

\section{The vector boson self-energy in the non-linear $\s$-model\label{c2}}

\subsection{The model\label{22} }

Our starting point is the action of the 3-dimensional $SU(2)$ Higgs model,
which is given by

\beq
S = \int d^3x \,\mbox{Tr} \left[ {1\over 2} W_{\m\n} W_{\m\n} + \left(D_\m\Phi\right)^\dagger D_\m\Phi + \m^2\Phi^\dagger\Phi + 2\l \left(\Phi^\dagger\Phi\right)^2\right]\, ,\label{linhix}
\eeq

\noindent
with

\beq
\Phi = {1\over 2}\left(\s+i\vec{\p}\cdot\vec{\t}\right), D_\m\Phi = \left(\partial - igW_\m\right)\Phi, W_\m = {1\over 2}\vec{\t}\cdot\vec{W}_\m \, . 
\eeq
 
Here $\vec{W}_\m$ is the vector field, $\s$ is the Higgs field, $\vec{\p}$ is 
the Goldstone boson field and $\vec{\t}$ the triplet of Pauli matrices.
To obtain the non-linear $\s$-model, one eliminates one degree of freedom by the
constraint
\beq
\s^2 = v^2 -\pi^2\, ,
\eeq
\noindent and takes the limit $\l,\m \to \infty$.
Setting $m={g^2 v^2\over 4}$, one then arrives at the following 
Lagrangian,

\beqa
\cl &=& {1\over 4} \vec{W}_{\m\n}\vec{W}_{\m\n} + {1\over 2\xi} \left(\partial_\m\vec{W}_\m\right)^2 + {1\over 2 }m^2 \vec{W}_\m^2\NO\\[1ex]
&& + {1\over 2}\left(\partial_\m\vec{\pi}\right)^2
+{\xi\over 2}m^2\vec{\pi}^2+{g\over 2}\left(\vec{W}_\mu \times \vec{\pi}\right)\NO\\[1ex]
&&+\partial_\m\vec{c^\ast}\partial_\m\vec{c} + \xi m^2 \vec{c^\ast}\vec{c}+g\partial_\m\vec{c^\ast}\cdot\left(\vec{W}_\m\times\vec{c}\right)\NO\\[1ex]
&&+\xi{g\over 2}m \vec{c^\ast}\cdot\left(\vec{\pi}\times \vec{c}\right)
-\xi{g^2\over 8}\vec{\pi}^2 \vec{c^\ast}\vec{c}\NO\\[1ex]
&&+{g^2\over 8}{\left(\vec{\pi}\partial_\m\vec{\pi}\right)^2\over m^2}
-{g^2\over 4}\vec{W}_\m\cdot\vec{\pi} {\vec{\pi}\cdot\partial_\m\vec{\pi}\over m}+{g^2\over 8}\vec{W}_\m\cdot\partial_\m\vec{\pi} {\vec{\pi}^2\over m}\, .\NO\\[1ex]
\label{noli}
\eeqa

Note that we have neglected all higher-dimensional operators which do not contribute
to the two-loop self-energy.
In  unitary gauge the unphysical degrees of freedom decouple and one is left with a massive Yang-Mills theory,

\beq
{\cal L} = {1\over 4} F^a_{\mu\nu} F^a_{\mu\nu} + {1\over 2} m^2 
W^a_{\mu} W^a_{\mu} 
\;,\label{rrr}
\eeq
with
\beq
F^a_{\mu\nu} = \partial_{\m}W^a_{\n}-\partial_{\n}W^a_{\m}+ g 
\e^{abc} W^b_{\m} W^c_{\n}
\;.\label{ddd}
\eeq

The advantage of the unitary gauge is that only a minimal amount of diagrams has to be calculated.
The one-loop self-energy in unitary and renormalizable gauges can be found
in 
\cite{BP,JP1}.
On mass-shell they coincide due to the BRS-invariance of the Lagrangian.

In the non-linear $\s$-model
there are non-renormalizable vertices. At the one-loop level, no problem 
concerning 
renormalization arises, as the non-renormalizable couplings 
do not contribute to this loop order.
Moreover, in dimensional regularization,
all one-loop integrals are finite in 3 dimensions.
To two loops the situation is more difficult.

\subsection{Two-loop self-energy in massive Yang-Mills theory}

\begin{figure}
\begin{center}
\psfig{file=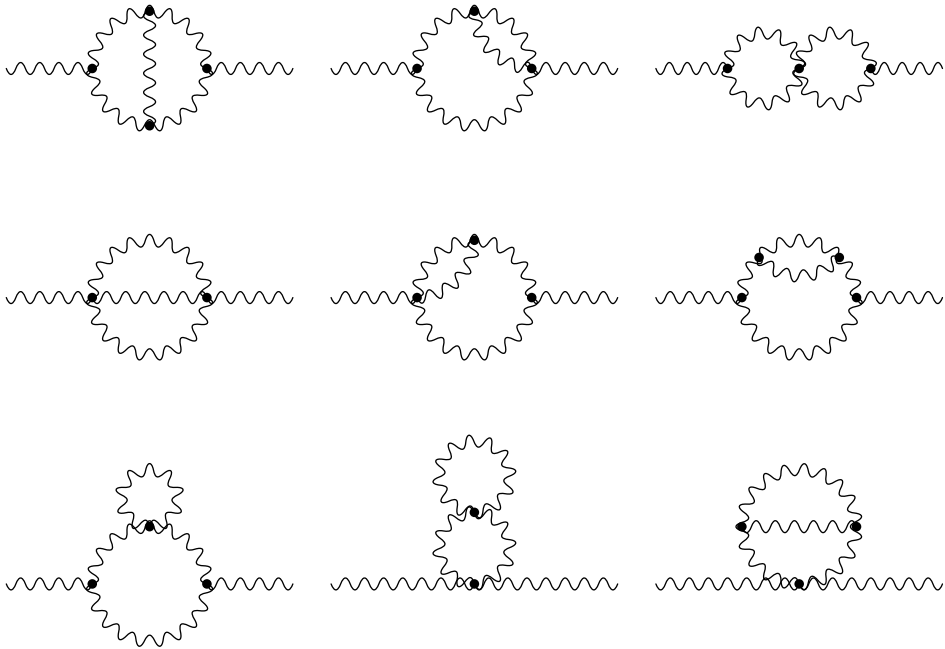}
\caption{ \it Diagrams contributing to the two-loop gap equation in the unitary gauge\label{fip}}
\end{center}
\end{figure}

In unitary gauge, only 9 two-loop
diagrams have to be evaluated for the non-linear $\s$-model. They are depicted in fig.~\ref{fip}.
\input{psfig}

As all propagators are massive and the external momentum does not vanish,
the reduction of the scalar integrals to basic integrals with no momenta in the
numerators turns out to be the most difficult step in the calculation.
 For  propagator type integrals this task has been achieved only recently
by Tarasov \cite{Tar}. Using his recurrence relations it is possible 
to reduce the self-energy integrals to a small set of 
linearly independent basic integrals.
For the first time, this  method  achieves  a  complete reduction
and stays on an algebraic level as far as possible.
Since the recurrence relations are in some cases quite involved,
they have to be implemented into a FORM package \cite{verm}. 
In  unitary gauge, the situation is even more complex due to
the high powers of momenta in the numerator. 
A peculiarity of the unitary gauge is that
the limit $\xi \to \infty$ must be performed before divergent integrals are evaluated \cite{jak}. Otherwise, one would get an infinite result for the self-energy.

The reduction program in FORM yields for the sum of the transverse parts

\beqa
 {1\over g^4} \Pi^{2-loop}_T(p^2) &=&
\left( {63\over 4}m^4 -{111\over 8}p^2m^2-{67\over 16} p^4
-{33\over 32} {p^6\over m^2} +{1\over 16} {p^8\over m^4}\right) F(m,m,m,m,m)
\NO\\[1ex]
&& + \left( -{63\over 2}m^2 -{113\over 8}p^2-{27\over 16} {p^4\over m^2}
+{109\over 64} {p^6\over m^4}\right) V(m,m,m,m)\NO\\[1ex]
&& + \left( -{189\over 2}{m^4\over p^2}
+{237\over 4}m^2 +{12257\over 80}p^2-{21\over 8} {p^4\over m^2}
-{167\over 80} {p^6\over m^4}\right) I_{211}(m,m,m)\NO\\[1ex]
&& + \left( {63\over 4}{m^2\over p^2}
-{111\over 8} -{159\over 16}{p^2\over m^2}+{463\over 192} {p^4\over m^4}
-{1\over 60} {p^6\over m^6}\right.\NO\\[1ex]
&&\,\,\,\,\left. 
-{387\over 4}{m^2\over p^2}\e
+{903\over 8}\e +{1597\over 20}{p^2\over m^2}\e -{149\over 15} {p^4\over m^4}\e
-{1\over 50} {p^6\over m^6}\e \right)
 I_{111}(p^2)(m,m,m)\NO\\[1ex]
&& + \left( -{63\over 4}{m^2\over p^2}
+{111\over 8} +{159\over 16}{p^2\over m^2}-{117\over 64} {p^4\over m^4}
\right.\NO\\[1ex]
&&\,\,\,\,\left. 
+{135\over 4}{m^2\over p^2}\e
-{195\over 8}\e -{207\over 8}{p^2\over m^2}\e -{3\over 4} {p^4\over m^4}\e
 \right)
 I_{111}(0)(m,m,m)\NO\\[1ex]
&& + \left( {37\over 4}m^2
-{5\over 2}p^2-{387\over 32} {p^4\over m^2}
-{1\over 32} {p^6\over m^4}
+{35\over 128} {p^8\over m^6}- {1\over 64} {p^{10}\over m^8}
\right) B^2(p^2,m^2,m^2) \NO\\[1ex]
&& + \left( {23\over 2}
-{151\over 8}{p^2\over m^2}-{57\over 8} {p^4\over m^4}
+{1\over 4} {p^6\over m^6}
+{1\over 16} {p^8\over m^8}
\right) B(p^2,m^2,m^2)A(m^2)\NO\\[1ex]
&& + \left( 
-{25\over 8}{1\over m^2}+{7\over 8} {p^2\over m^4}
+{87\over 160} {p^4\over m^6}
-{1\over 16} {p^6\over m^8}
\right) A(m^2) A(m^2)\, .\label{p2os}
\eeqa

%If one expands the basic integrals in \cite{raj} around $p^2=0$ 
%and takes $\e\to 0$, the terms $\sim {1\over p^2}$ in eq.~(\ref{p2os})
% cancel, leaving
%a well-defined limit $p^2\to 0$ as required.
\noindent The sum of the longitudinal parts 
adds up to 0 for all external momenta $p$,
which is a nice check of the calculation.
\input{psfig}
\begin{figure}
\begin{center}
\psfig{file=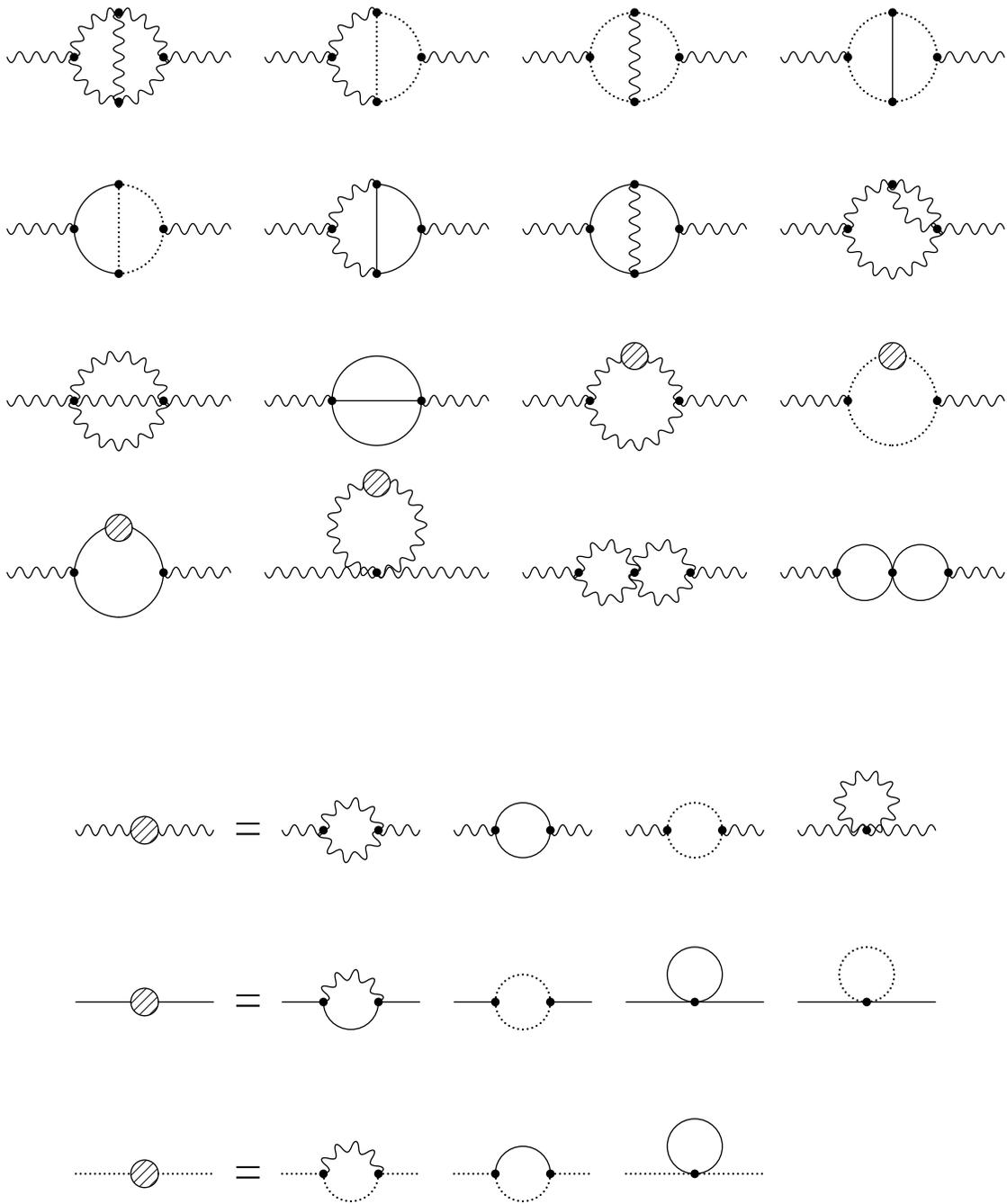}
\caption{ \it Generic two-loop self-energy diagrams in the non-linear $\s$-model \label{fid}}
\end{center}
\end{figure}

In 3 dimensions,
$I(p^2)(m,m,m)$ and $I(0)(m,m,m)$ are logarithmically UV-divergent,
whereas all other basic integrals are finite in dimensional regularization.
In $d=3-2\e$,  these two integrals 
 exhibit the following behaviour for small $\e$,

\beq
I(p^2)(m,m,m) = I(0)(m,m,m) = {1\over 64 \pi^2 \e} + \mbox{finite}\, ,
\eeq
leading to poles in the self-energy,

\beq
\Pi_T^{2-loop} = \left({7\over 12} {p^4\over m^4}-{1\over 60} {p^6\over m^6}\right)
 {1\over 64 \pi^2 \e} + \mbox{finite}\, ,
\eeq
which cannot be dealt with by a mass or wave function renormalization.
As we will see in the next section, this is
due to the bad high-energy behaviour of the propagator in unitary gauge.
A similar problem arises already at one
loop, if one uses cutoff-regularization.
Calculations of counter-terms
cannot be done in unitary gauge.
However, if one is interested in finite parts of gauge-invariant quantities
like poles in propagators, the unitary gauge provides
a convenient short-cut of the calculation.

\subsection{Two-loop calculation in Feynman gauge}

The diagrams which have to be evaluated in Feynman gauge
 are depicted in figure \ref{fid}.
In the  sum of the transverse parts of the generic two-loop diagrams  
$\Pi_T^{2-loop}$, 
the coefficients in front of the products of one-loop basic integrals 
do not coincide in unitary and in Feynman gauge.

However, this gauge-dependence combined with the gauge-dependence in 
${\partial \over \partial p^2}\Pi_T^{1-loop}$ leads to a gauge-invariant result for
$\Pi_T(p^2=-m^2)\left(1+{\partial \over \partial p^2}\Pi_T(p^2=-m^2)\right)$ 
at the  two-loop level.
This quantity is nothing but the two-loop 
 pole of the propagator, which should be 
gauge-invariant in BRS-symmetric theories \cite{rebhan}.
%The gauge-dependent derivative 
%of the one-loop self-energy is,
%\beq
%{\partial \over \partial p^2}\Pi_T^{1-loop}(p^2= - m^2) =  {1\over 8\pi}\l%eft({33\over 4} - {21\over 16}
%\mbox{ln} 3 +(\xi -{1\over 4}) \mbox{ln} {2\sqrt{\xi}+1\over 2\sqrt{\xi}-1%}   -3\sqrt{\xi } \right)l{g^2\over m} 
%\;,\; \xi > {1\over 4}\; 
%\;.\label{hai}
%\eeq
The
preceding statements can be checked using the result of the last section and the
sum of  the generic two-loop diagrams in Feynman gauge written
in  appendix \ref{ap2}. We
obtain the same position for the pole of the propagator as in unitary gauge.

The Feynman gauge is a renormalizable gauge.
Collecting the coefficients of 
$I(p^2)(m,m,m)$ and $I(0)(m,m,m)$, we get 
the following poles in $\e$ for the self-energy

\beq
{1\over g^4}\Pi_{T,\xi = 1}^{2-loop}(p^2) = \left({7\over 12}-{1\over 60} {p^2\over m^2}
\right) {1\over 64 \pi ^2 \e} + \mbox{finite}
\;.\label{renn}
\eeq

According to eq.~(\ref{renn}), a mass and 
wave function renormalization  removes the infinities
in the two-loop self-energy. This is also suggested by naive power counting.

\section {The Higgs and vector boson self-energy in the $SU(2)$ Higgs model\label{c5}}
\subsection{The model\label{21}}

Consider now the 
 3-dimensional $SU(2)$
Higgs model of eq.~(\ref{linhix}).
Varying $\m^2/g^4$ one expects a phase transition which is
of first order for sufficiently small values of $\l/g^2$.

We change parameters according to

\beq
\m ^2 = {1\over 2} M^2, \l = {g^2\over 8} {M^2\over m^2}\, ,
\eeq

\noindent  shift the Higgs field $\s$ around its classical
minimum  $\, \s = {2m\over g}+\s^\prime$, and 
 add an  $R_\xi$-gauge fixing term
${\cal L}_{GF} = {1\over 2\xi}\left(\partial_\m W^a_\m - \xi m \pi^a\right) $ and the corresponding ghost terms in the usual way.
The  resulting Lagrangian reads,

\beqa
\cl_R &=& {1\over 4} \vec{W}_{\m\n}\vec{W}_{\m\n} + {1\over 2\xi} \left(\partial_\m\vec{W}_\m\right)^2 + {1\over 2 }m^2 \vec{W}_\m^2\NO\\[1ex]
&& +{1\over 2}\left(\partial_\m\s^\prime\right)^2 +{1\over 2}M^2 \s^{\prime 2}
+ {1\over 2}\left(\partial_\m\vec{\pi}\right)^2
+{\xi\over 2}m^2\vec{\pi}^2\NO\\[1ex]
&&+{g\over 2}m\s^\prime\vec{W}_\m^2 + {g\over 2}\vec{W}_\m\cdot \left(\vec{\pi}\partial_\m\s^\prime-\s^\prime\partial_\m\vec{\pi}\right)+{g\over 2}\left(\vec{W}_\mu \times \vec{\pi}\right) \cdot \partial_\m\vec{\pi}
\NO\\[1ex]
&&+{g^2\over 8}\vec{W}^2_\m \left(\s^{\prime 2}+\vec{\pi}^2\right)
+{g\over 4} {M^2\over m}\s^\prime\left(\s^{\prime 2}+\vec{\pi}^2\right)
+{g^2\over 32} {M^2\over m^2}\left(\s^{\prime 2}+\vec{\pi}^2\right)^2\NO\\[1ex]
&&+\partial_\m\vec{c^\ast}\partial_\m\vec{c} + \xi m^2 \vec{c^\ast}\vec{c}\NO\\[1ex]
&&+g\partial_\m\vec{c^\ast}\cdot\left(\vec{W}_\m\times\vec{c}\right)
+\xi{g\over 2}m\s^\prime \vec{c^\ast}\vec{c}+\xi{g\over 2}m \vec{c^\ast}\cdot\left(\vec{\pi}\times \vec{c}\right)
\, .\label{rtre}
\eeqa

\noindent The one-loop results for the self-energies in $R_\xi$-gauge can be found in \cite{BP}.
The on-shell self-energies coincide for all gauges.

\subsection{Two-loop self-energy for the Higgs field}

The generic two-loop Higgs 
self-energy 
diagrams are depicted in appendix \ref{ap4}.
The sum of these diagrams is evaluated
 on mass-shell, $p^2 = - M^2$.
The lengthy expressions resulting from the reduction to
basic integrals can also be found in appendix \ref{ap4} in 
Feynman gauge and in unitary gauge.
For the sum of the generic two-loop diagrams $\Sigma^{2-loop}$,
the unitary gauge result differs from the result in Feynman gauge
only by products of one-loop integrals.
Quantitatively,

\beqa
&&\Sigma_{\xi = 1}^{2-loop}(p^2=-M^2)-\Sigma_{\xi = \infty}^{2-loop}(p^2=-M^2)
= \NO\\[1ex]
&&\Sigma_{\xi = 1,\infty }^{1-loop} (p^2=-M^2)  \left({\partial\over \partial p^2} \Sigma_{\xi = 1 }^{1-loop} (p^2=-M^2) - {\partial\over \partial p^2} \Sigma_{\xi = \infty }^{1-loop} (p^2=-M^2)\right)\, .\label{diif}
\eeqa

This ensures that neglecting the resummation counter-terms,
the pole of the Higgs boson propagator is gauge parameter independent
to two loops.
The underlying reason for this powerful check of the calculation is 
the BRS-invariance of the linear model.
Eq.~(\ref{diif}) can be verified using the expressions
 in appendix \ref{ap4} and the
one-loop results in \cite{BP} and section \ref{21}.

Concerning renormalization, eq.~(\ref{hiegs}) leads to
 the following pole structure in $\e$
for the Higgs self-energy (keeping the external momentum $p^2$ arbitrary),

\beq
\Sigma^{2-loop}_{\xi < \infty}(p^2) = \left( {51\over 8} + {9\over 4} {M^2\over m^2} - {3\over 8} {M^4\over m^4}\right) {1\over 64\pi^2\e} + \mbox{finite}.\label{cth}
\eeq

As expected in the super-renormalizable 3-dimensional
Higgs model, no
wave function renormalization is necessary.
For the Higgs field, we have to add only 
a mass renormalization counter-term.

\subsection{Two-loop self-energy for the vector field}

The analogous calculation is performed for the vector boson field.
The sum of the generic two-loop diagrams shown in appendix \ref{ap4}
 is evaluated in Feynman and unitary gauge
in $d=3-2\e$ on mass-shell, $p^2 = - m^2$ (note, that we only draw those Feynman diagrams,
whose transverse part is non-zero).

A relation similar to eq.~(\ref{diif}) holds for the transverse
two-loop vector field self-energy.
As for the Higgs field, it can be verified,
 that the pole of the transverse part of the
vector boson propagator is
gauge-invariant to two-loop order.
In dimensional regularization, the  divergent terms in the vector self-energy (evaluated  for arbitary external momentum $p^2$) can be obtained from eq.~(\ref{veict}),

\beq
\Pi_T^{2-loop}(p^2) = \left( {51\over 8}{m^2\over M^2} + {9\over 4} - {3\over 8} {M^2\over m^2}\right) {1\over 64\pi^2\e} + \mbox{finite}\, .\label{ctv}
\eeq

As for the Higgs field,  no wave function 
renormalization  is needed for the vector field.
There is only a renormalization of the vacuum expectation value.
Comparing eq.~(\ref{ctv}) with eq.~(\ref{cth}), one can see a simple relation 
between the 
divergent terms:
the coefficients of the divergent terms
differ only by the  factor ${m^2\over M^2}$ 
(Ward-identity).

%The proved gauge-invariance of the poles of the 
% Higgs  and vector boson propagator 
%to two-loops constitutes a powerful test for our FORM package.

%\input{c3p}

\section {Two-loop gap equation in the $SU(2)$ Higgs model\label{c6}}

In the high-temperature phase of
the standard model a naive perturbative expansion
with a vanishing vector boson mass leads to severe
infrared divergences in the magnetic sector of the theory \cite{vier}.
Introducing a non-vanishing mass which acts as an infrared cut-off 
can cure these problems. The symmetric phase  is expected to be governed by  
non-perturbative effects whose size is determined by
the magnetic screening length, which is the inverse of 
the
 magnetic mass.
In an apparently massless 3-dimensional
Yang-Mills theory, 
 the gauge coupling $g^2$ carries the dimension of mass, thereby
providing a natural mass scale.

The propagator mass of the vector boson can be calculated analytically using gap equations \cite{BP}.
In this chapter, we investigate two-loop effects on the gap equation
in the 3-dimensional Higgs model.
%Many more diagrams have to be evaluated than in the non-linear case.

\subsection {Gap equations and resummation\label{pop}}

We again consider  the model defined by  eq.~(\ref{linhix}).
We are interested in the Higgs and vector boson masses in both phases which
determine
the exponential fall-off of the corresponding two-point functions at large
separation $|x-y|$,

\beqa
\langle \s(x) \s(y)\rangle &\sim& e^{-M|x-y|}\, ,\NO\\[1ex]
\langle W_\m(x) W_\m(y)\rangle &\sim& e^{-m|x-y|}\, .
\eeqa

In the Higgs phase, these 2-point functions can be evaluated in perturbation theory. The masses $m$ and $M$ are given by the gauge-independent poles of the corresponding propagators in momentum space.
In eq.~(\ref{linhix}) we  shift the Higgs field $\s$ around its vacuum expectation value $v,\, \s = v+\s^\prime$,  add an  $R_\xi$-gauge fixing term and the corresponding ghost terms in the usual way.
This yields the following masses for the vector boson and the Higgs field, 

\beq
m_0^2 = {g^2\over 4}v^2,\,  M_0^2 = \m^2+3\l v^2\, .
\eeq
The ghost and Goldstone boson mass is given by $\sqrt{\xi} m_0$.

In order to extract a non-vanishing mass in the symmetric phase, where  
in ordinary perturbation theory $v=0$, we add and subtract a mass-term.
The tree-level masses $m_0^2$ and $M_0^2$ 
are expressed as 

\beq
m_0^2 = m^2-\d m^2\, ,\,  M_0^2 = M^2-\d M^2\, ,
\eeq

\noindent where $m$ and $M$ enter the propagators of the loop expansion, and $\d m^2$ and
$\d M^2$ are treated perturbatively as counter-terms.
For a
gauge-invariant  one-loop gap equation it is necessary and
sufficient to have a BRS-invariant resummed tree-level 
action.
This requires a suitable resummation of the ghost and Goldstone
boson mass as well as of the following  vertices,

\beq
{g^2 v\over 2} = gm - \d V^g,\,
\l v = {gM\over 4m} -\d V^{\l v},\,
\l = {g^2M^2\over 8m^2} - \d V^\l .
\eeq
The  resulting Lagrangian reads \cite{BP},

\beqa
\cl &=&\cl_R + \cl_1 + \cl_0\, ,\NO\\[1ex]
\cl_1 &=& -\d m^2 \left( {1\over 2}\vec{W}^2_\m + {\xi\over 2}\vec{\pi}^2 + \xi \vec{c^\ast} \vec{c}\right) - {1\over 2}\d M^2\s^{\prime 2} +{1\over 2}\left(\m^2+\l  v^2\right)\vec{\pi}^2 \NO\\[1ex]
&&+v\left(\m^2+\l v^2\right)\s^\prime - {1\over 2}\d V^g \left(\s^\prime \vec{W}_\m^2 + \xi\s^\prime \vec{c^\ast}\vec{c}+\xi \vec{c^\ast}\cdot\left(\vec{\pi}\times \vec{c}\right)
\right)\NO\\[1ex]
&&-\d V^{\l v}\s^\prime\left(\s^{\prime 2}+\vec{\pi}^2\right)
-{1\over 4}\d V^\l \left(\s^{\prime 2}+\vec{\pi}^2\right)^2\, ,\NO\\[1ex]
\cl_0 &=& {1\over 2}\mu^2 v^2 +{1\over 4}\l v^4\, ,
\eeqa
where $\cl_R$ equals the one in eq.~(\ref{rtre}),

In resummed perturbation theory, the vertices defined by $\cl_1$ 
are treated as counter-terms.
%The propagators for the vector boson,
%Goldstone boson, ghost and Higgs boson, respectively, are easily obtained, 
%\beqa
%D^{ab}_{\m\n}(p) &=& \d_{ab}{1\over p^2+m^2}\left( \d_{\m\n}+(\xi -1)
%{p_\m p_\n\over p^2+\xi m^2}\right)\, ,\NO\\[1ex]
%\bigtriangleup_\pi^{ab}(p) &=& \bigtriangleup_c^{ab}(p) = {\d_{ab}\over p^2+\x%i m^2}
%\, ,\NO\\[1ex]
%\bigtriangleup_\s^{ab}(p) &=& {1\over p^2+M^2}\, .
%\eeqa
The coupled set of gap equations for the poles of the  Higgs and vector propagator
then reads,

\beqa
{\Pi_T(p^2=-m^2)\over 1 - {\partial\over \partial p^2} \Pi_T (p^2 = -m^2)} &=& 0\, ,\NO\\[1ex]
{ \Sigma(p^2=-M^2)\over 1 - {\partial\over \partial p^2} \Sigma (p^2 = -M^2)}&=& 0\, ,\NO\\[1ex]
\langle \s^\prime \rangle &=& 0\, ,\label{gapset}
\eeqa
where $\Sigma $ is the one-loop Higgs boson self-energy and $\Pi_T$ is the transverse part of the vacuum polarization tensor.
In resummed perturbation theory, one expands
eq.~(\ref{gapset}) to the desired order and solves the set of gap equations
for $m$. 
In theories with a BRS-symmetry the position of the pole of the propagator
and therefore the first two  eqs. in (\ref{gapset}) are
 gauge-independent on mass-shell \cite{rebhan}.
The self-energy itself is not gauge-invariant on mass-shell except at the one-loop level. Only to one-loop, the denominators in the LHS of eq.~(\ref{gapset}) can be
neglected.

%\beq
%\Pi^{ab}_{\m\n}(p) = \d_{ab}\left[\left(\d_{\m\n}-{p_\m p_\n\over p^2}\right)
%\Pi_T(p^2) + {p_\m p_\n\over p^2}\Pi_L(p^2)\right]\, .
%\eeq

To one loop, the third equation of (\ref{gapset}), which
determines the vacuum expectation value $v$ of the Higgs field
self-consistently, is not gauge parameter 
independent since $v$ is no physical observable. 
On the other hand, the masses obtained form the gap equations
(\ref{gapset}) must be gauge independent.
The weak gauge dependence induced by the gauge dependence of $v$ therefore
has to be cancelled
by higher order contributions.

Details of the one-loop calculation in renormalizable gauges
 and the solutions of the gap equations
in the linear Higgs model in Landau gauge can be found in \cite{BP}.
The main result is that deeply in the symmetric phase, the value for the gap
mass is approximately the same as the one obtained in a non-linear $\sigma $
model, which requires the evaluation of much less diagrams. The analytical 
result for the one-loop gap mass  in the non-linear $\sigma$-model is,

\beq
m =  {1\over 16\pi}({63\over 4}\mbox{ln} 3 - 3)\, g^2 \simeq 0.28\, g^2 \, .
\eeq

%\noindent In the Higgs model, it is approximately the same.

\subsection {Two-loop gap equation in the non-linear $\s$-model\label{poprr}}

The two-loop calculation of the gap equation in the non-linear $\s$-model using the self-energies evaluated in chapter \ref{c2}
can be found in \cite{frenk}.
 It is a crucial test for the consistency 
of the whole
approach since the loop expansion  does not correspond to an  expansion in a small
parameter ${g^2\over m}$.
 Nevertheless, this does not exclude
that the one-loop result provides a  reasonable approximation
for the true mass gap. This is  a question of numerical factors.
The two-loop gap equation is quadratic in $m$, whereas at one loop it is
linear.
The existence of a positive solution is therefore 
a  non-trivial check for the method.
The results for the gap mass are listed in table \ref{taby1}.

\begin{table}
\renewcommand{\arraystretch}{2}
\setlength{\arrayrulewidth}{0.3mm}

\begin{center}
\begin{tabular}{|c|c|c|c|}

 \hline
${\mu_{\overline{MS}}\over m}$ & $0.3$ & $1$ & $3$  \\
\hline
${m\over g^2}, \xi = 1,\infty $ & $0.343$ & $0.335$ & $0.327$ \\
\hline
${m\over g^2}, \xi = 2 $ & $0.345$ & $0.336$ & $0.328$ \\
\hline
${m\over g^2}, \xi = 10 $ & $0.350$ & $0.342$ & $0.334$ \\
\hline  
 \end{tabular}
\caption {\label{taby1} \it Solutions of the two-loop gap equation}
 \end{center}
\end{table}

The calculation in the non-linear model shows, that 
the two-loop correction to the one-loop gap mass is only $15-20 \%$.
The dependence on the gauge parameter $\xi$ and the renormalization scale $\mu_{\overline{MS}}$
is very small numerically. For a more detailed discussion see \cite{frenk}.
The two-loop gap mass is in good agreement with the results form 
the other one-loop calculations \cite{JP1,AN,C} and in perfect aggrement with the lattice result in \cite{kar}.
To judge the significance of the result in the non-linear case, it is crucial
to perform the whole calculation in the Higgs model, which is 
super-renormalizable.

\subsection{Two-loop gap equation in the Higgs model}

In the two-loop calculation, we will not solve the
complete set of three gap equations, but restrict ourselves to the
first of eqs.~(\ref{gapset}), the gap equation for the vector boson mass.
For different values of
$z=M/m$, we insert the corresponding $\m$ and $v$ from 
the one-loop
solution and then solve the equation for $m$.
We also investigate the dependence of $m$ on varying 
$\m$ and $v$ around the one-loop value.

The two-loop gap equation for the vector boson
 is gauge parameter dependent.
First, as in the one-loop case, this is caused by a $\xi$-dependent $v$.
Second, as in the two-loop case in the non-linear $\s$-model, it is
due to the one-loop (resummation) counter-term diagrams.
We perform the two-loop calculation in
the linear Higgs model in Feynman gauge,
in contrast to the one-loop calculation in \cite{BP}, where  Landau gauge, $\xi = 0$, is used.
For a suitable comparison of one- and two-loop results,
 we first solve the one-loop gap equations in Feynman gauge.

Table \ref{tab3} shows the one-loop solutions in Feynman gauge
for
$\mu , v $ and $m$ for different values of $z$, with ${\lambda\over g^2} = {1\over 8}$.
From the treatment in Landau gauge in \cite{BP} we see, that
$1 \le z \le 2$  is a reasonable choice for the symmetric phase.
$z \ge 2$ is forbidden, since in this case the Higgs boson can decay into
two vector bosons.
As a consequence of this, there will be poles in the two-loop result for the self-energy for $M=2m$.
The one-loop gap equation for $M$ is complex for $z > 2$.

%We perform the two-loop calculation in
%the linear Higgs model in Feynman gauge,
%in contrast to the one-loop calculation in \cite{BP}, where  Landau gauge is u%sed.
%For a suitable comparison of one- and two-loop results,
% we first solve the one-loop gap equations in Feynman gauge.
%In   table \ref{tab3}, we calculate the one-loop solutions
%for
%$\mu , v $ and $m$ for different values of $z$, with ${\lambda\over g^2} = {1\%over 8}$.
%We use  Feynman gauge, $\xi = 1$.
%The equations are listed in \cite{BP}.
%From the treatment in Landau gauge in \cite{BP} we see, that
%$1 \le z \le 2$  is a reasonable choice for the symmetric phase.
%$z \ge 2$ is forbidden, since in this case the Higgs boson can decay into
%two vector bosons.
%As a consequence of this, there will be poles in the two-loop result for the s%elf-energy for $M=2m$.
%The one-loop gap equation for $M$ is complex for $z > 2$.

\begin{table}
\renewcommand{\arraystretch}{2}
\setlength{\arrayrulewidth}{0.3mm}

\begin{center}
\begin{tabular}{|c|c|c|c|c|c|}

 \hline
$z$ & $1$ & $1.2$ & $1.4$ & $1.6$ & $1.8$ \\
\hline
$ {\mu ^2\over g^4}$ & $0.111$ & $0.153$ & $0.208$ & $0.277$ & $0.367$\\
\hline
$ {v\over g} $ & $0.159$ & $0.162$ & $0.160$ & $0.155$ & $0.148$\\
\hline
$ {m\over g^2} $ & $0.226$ & $0.231$ & $0.234$ & $0.236$ & $0.237$\\
\hline  
 \end{tabular}
\end{center}
\caption{\label{tab3} \it Solutions of the one-loop gap equation in Feynman gauge}

\end{table}

It can be seen that there is a constant value for the vector boson
mass deeply in the symmetric phase.

We aim at a solution of  eq.~(\ref{gapset}) to two-loop order.
 Since  solving  the
 complete set of three gap equations (\ref{gapset})
 would be unnecessarily complicated, we will use the following short-cut.
We look at first equation in eq.~(\ref{gapset}), the gap equation
for the vector boson, for different values of $z$, which are typical for the symmetric phase according to the one-loop calculations.
For $\m ^2$ and $v$ we will insert the corresponding one-loop results
from table \ref{tab3}.
As already explained, the gap equation is gauge parameter dependent.
We restrict the discussion to  Feynman gauge.

\begin{figure}
\begin{center}
\psfig{file=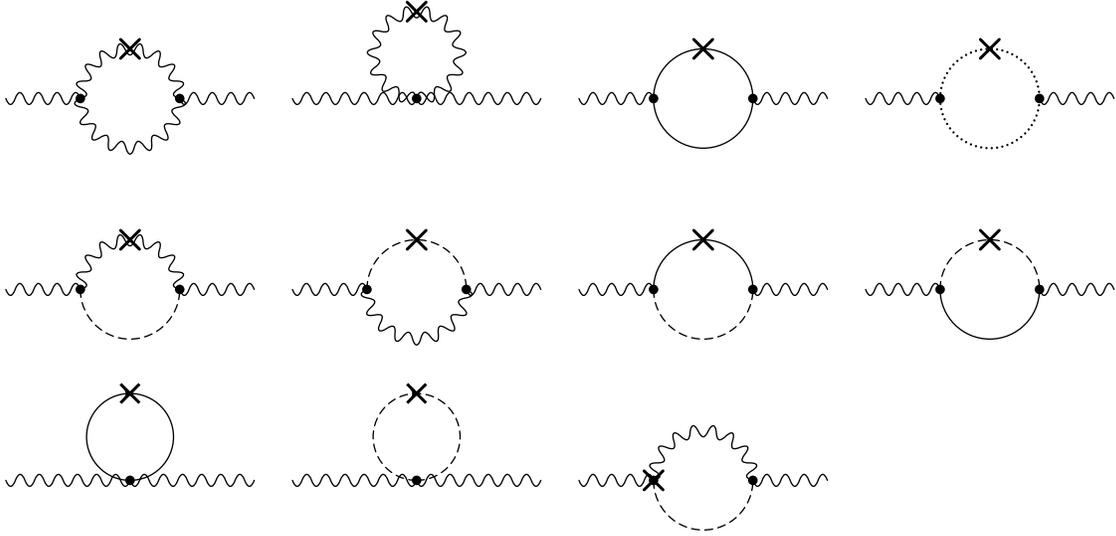}
\caption{ \it One-loop CT diagrams contibuting to the gap equation\label{ytu}}
\end{center}
\end{figure}

In setting up the vector field gap equation we have to insert the
 third equation
of (\ref{gapset}).
This condition reduces the amount of one-loop counter-term and generic
two-loop self-energy diagrams which contribute to the first equation of (\ref{gapset}).
We can leave out all the two-loop 
diagrams involving tadpoles as well as the one-loop counter-term
diagrams which contain the scalar one-point function.
The remaining one-loop  diagrams with resummation counter-terms contributing to the first equation of (\ref{gapset})
are depicted in fig.~\ref{ytu}. Their on-shell value is given by

 \input{psfig}

\beqa 
&&\Pi_T^{1-loop-CT} (p^2 = - m^2) =
\d m^2 \left[ - {1\over 64 \pi m} \left( 50 + 2 {M\over m} + {M^2\over m^2}\right)\right.\NO\\[1ex]
&& + {1\over 64\pi (2m+M)} \left( {M^2\over m^2} + 8 {m^2\over M^2} - 4 \right)
 + \left.{9\over 16\pi m} \mbox{ln} 3 + {1\over 32\pi} {M^2\over m^3} \mbox{ln}
\left( 1+2{m\over M}\right) \right]\NO\\[1ex]
&&+ \d M^2 \left[  {1\over 64 \pi } \left( 3 {M\over m^2} - {2 \over m} - {2\over M}\right)\right. + {1\over 64\pi (2m+M)} \left( {M^3\over m^3} - 4 {M\over m} + 8 {m\over M} \right)\NO\\[1ex]
&& + \left.{1\over 32\pi m} \left(2 - {M^2\over m^2}\right) \mbox{ln}
\left( 1+2{m\over M}\right) \right] + {\d V \over 4\pi }  \mbox{ln}
\left( 1+2{m\over M}\right) \NO\\[1ex]
&& - \left( \m ^2+\l v^2\right)\left[  {1\over 64 \pi m } \left( 2  + 2 {M \over m} + {M^2\over m^2}\right)\right. + {1\over 64\pi (2m+M)} \left(  4  -  {M^2\over m^2} \right)\NO\\[1ex]
&& - \left.{1\over 32\pi }  {M^2\over m^3} \mbox{ln}
\left( 1+2{m\over M}\right) - {1\over 16\pi m } \mbox{ln} 3\right]\, .
\eeqa

With all the quantities evaluated above, we are now in the position
to discuss the two-loop gap equation for the vector boson field,

\beqa
&& m^2 -  {g^2\over 4} v^2 + \Pi_T^{1-loop} (p^2=-m^2)\NO\\[1ex]
&&+ \Pi_T^{2-loop} (p^2=-m^2) + \Pi_T^{1-loop-CT} (p^2=-m^2)\NO\\[1ex]
&&+{\partial\over \partial p^2} \Pi_T^{1-loop}(p^2=-m^2) \left(\Pi_T^{1-loop} (p^2=-m^2)+ m^2 -  {g^2\over 4} v^2\right) = 0\label{ghap}\, .
\eeqa

The gap equation is investigated 
for different $z$, with $1\le z\le 2$.
We choose $z=1,1.2,1.4,1.6,1.8$.
%For these values of $z$ the master integral
%$F(m_1,m_2,m_3,m_4,m_5), m_i=m,M$, is evaluated in appendix \ref{ap5} .
As in the non-linear $\s$-model, we work in the ${\overline{MS}}$-scheme.
It turns out, that the coefficient in front of the
$\mu _{\overline{MS}}$-dependent terms is negligibly small.
Therefore, we set $\mu _{\overline{MS}} = m$ in what follows.

 The  solutions of eq.~(\ref{ghap}) for different values of z 
are listed in table \ref{tab4}. They are compared with the one-loop result in 
Feynman gauge.

\begin{table}
\renewcommand{\arraystretch}{2}
\setlength{\arrayrulewidth}{0.3mm}

\begin{center}
\begin{tabular}{|c|c|c|c|c|c|}

 \hline
$z$ & $1$ & $1.2$ & $1.4$ & $1.6$ & $1.8$ \\
\hline
$ {m\over g^2} $ (one-loop) & $0.226$ & $0.231$ & $0.234$ & $0.236$ & $0.237$\\
\hline  
$ {m\over g^2} $(two-loop) & $0.303$ & $0.307$ & $0.310$ & $0.309$ & $0.299$\\
\hline  
 \end{tabular}
\caption {\label{tab4}\it Solutions of the two-loop gap equation in Feynman gauge}
\end{center}
\end{table}

 For the scalar coupling $\l\over g^2$, we choose ${1\over 8}$.
For this value, 
a crossover behaviour was found for the  transition between the Higgs and the symmetric phase.
As
table \ref{tab4} shows,
the two-loop solutions exhibits a similar behaviour as the one-loop gap mass:
it is numerically nearly independent of the value of the Higgs mass.
Moreover, the   two-loop correction is of the same sign and
approximately of the same size as the
correction in the non-linear $\s$-model.
The $15-20\%$ difference between the numerical value of the 
one-loop gap mass in the non-linear and linear model in Feynman gauge
 still remains
at two loops.

In solving eq.~(\ref{ghap}), the one-loop values for $\mu^2$ and $v$
 are inserted
for each value of $z$ according to table  \ref{tab3}.
At two loops these values change, if one solves the 
set of three gap equations exactly.
To estimate this effect, we vary $\m ^2$ and $v$ around the 
one-loop solutions of table \ref{tab3} for a fixed value of
$z$ ($z=1.6$) and show that there is only a small numerical influence on 
the two-loop gap mass (see tables \ref{tab5} and \ref{tab6}).

\begin{table}
\renewcommand{\arraystretch}{2}
\setlength{\arrayrulewidth}{0.3mm}

\begin{center}
\begin{tabular}{|c|c|c|c|}

 \hline
$z=1.6, {\m ^2\over g^4}=0.277, {v\over g}= $ & $0.1$ & $0.155$ & $2$ \\
\hline
$ {m\over g^2} $ & $0.303$ & $0.309$ & $0.317$  \\
\hline  
 \end{tabular}
\caption{\label{tab5} \it Solutions of the two-loop gap equation for different ${v\over g}$}
\end{center}
\end{table}

\begin{table}
\renewcommand{\arraystretch}{2}
\setlength{\arrayrulewidth}{0.3mm}

\begin{center}
\begin{tabular}{|c|c|c|c|}

 \hline
$z=1.6, {v\over g}=0.155, {\m ^2\over g^4}=$ & $0.177$ & $0.277$ & $0.377$ \\
\hline
$ {m\over g^2}$  & $0.304$ & $0.309$ & $0.315$  \\
\hline  
 \end{tabular}
\end{center}
\caption{\label{tab6} \it Solutions of the two-loop gap equation for different ${\m ^2\over g^4}$}
\end{table}
           
The small numerical difference between the gap mass in the linear and
non-linear model as well as 
the approximate independence of the gap mass of the Higgs mass M 
 shows that the 
non-linear $\s$-model describes the infrared limit
of the linear Higgs model and of the
electroweak Standard Model at finite temperature 
to a very good approximation.

\section{Conclusions}

We have investigated two-loop effects on the propagator  
in 
the following 3-dimensional theories:
a resummed massive Yang-Mills theory,
 a resummed non-linear $\s$-model
in arbitrary gauge and a resummed
 $SU(2)$ Higgs model in unitary and Feynman gauge.
%The one-loop approaches have been summarized and some additional calculations 
%have been performed, in particular the  one-loop self-energies of the vector 
%and Higgs field
%in the Higgs model in unitary gauge.
% The one-loop self-energy
%for the vector boson in the non-linear $\s$-model has been recalculated in 
%cut-off regularisation.
%In dimensional regularization all one-loop self-energies are finite.
%In cut-off regularization, the calculation in renormalizable gauges shows, tha%t
% a mass renormalization counter-term has to be introduced into the non-linear %$\s$-model at the one-loop level.
%In unitary gauge, calculating to one-loop order with a  cut-off
%leads to divergences with  high powers of the external momentum which
%cannot be removed by a renormalization of the mass or the wave-function. This %is because the unitary gauge is a non-renormalizable
% gauge. 

The two-loop calculation of the transverse vector self-energy in the non-linear
$\s$-model in unitary gauge shows
 divergences with  high powers of the external momentum. They
cannot be removed by a renormalization of the mass or the wave-function.
This is because the unitary gauge is a non-renormalizable
gauge. 
In renormalizable gauges, a mass and wave-function renormalization are sufficient to get rid of the infinities.
To three-loop order, naive power counting suggests that a similar
problem also arises
 in renormalizable gauges.
This is due to the non-renormalizability of the non-linear $\s$-model.
In the linear Higgs model, however, we have seen that at the  two-loop level
 a mass renormalization
is sufficient in  Feynman gauge, as expected in a super-renormalizable theory in 3 dimensions.
In unitary gauge of the linear model, however, the problematic situation remains.

The two-loop self-energies for the Higgs and the vector field have then been
applied
to set up the gap equations for the vector boson mass in the symmetric phase
of the Higgs model.
The two-loop gap equation  in a resummed non-linear $\s$-model
was already discussed in our previous paper \cite{frenk}  and shows a real and
positive solution for the vector boson mass $m \simeq 0.34\, g^2$.
%This is clearly a non-trivial result.
%It is only $\sim 20\, \% $ larger than the one-loop result and
%in very good 
%agreement with the lattice simulations of the propagator mass
%in Landau gauge and with
%the results of other one-loop models.
%The gap equation still contains a weak gauge dependence and a logarithmic
%dependence on the renormalization scale.
%It has been shown that they are numerically unimportant.
%This suggests that the solution  constitutes
% a reliable approximation
%to the exact gluon propagator mass in an $SU(2)$ gauge theory.
The corresponding  calculation in the super-renormalizable Higgs model is crucial to judge the significance of a calculation in the non-renormalizable non-linear sigma model.
%The vector-boson and Higgs self-energy have been calculated on mass-shell
%in Feynman and unitary gauge and the corresponding
We have solved the gap equation for the vector boson mass
varying the Higgs mass.

In the symmetric phase, the result for the gap mass  $m \simeq 0.31\, g^2$
is almost independent of the Higgs mass.
This proves that the non-linear $\s$-model is a very good
approximation for infrared phenomena of the
linear Higgs model.
Moreover, the two-loop correction in  the linear model is of similar size
as in the non-linear model.

A vector boson mass $\simeq 0.31  g^2$ or $\simeq 0.34  g^2$  is not in contradiction with confinement. It is of the same size as the confinement scale given by the string tension \cite{tepi}. The connection of such a propagator mass 
to the heavier
glueball masses  $\sim {\cal O}(1) g^2$  \cite{ptw} 
  remains to be clarified.

%To judge the significance of a calculation in the non-renormalizable non-linea%r sigma model,  we have performed
% the same calculation in the Higgs model, which is super-renormalizable.
%The vector-boson and Higgs self-energy have been calculated on mass-shell
%in Feynman and unitary gauge and the corresponding
%gap equation for the vector boson mass has been solved
%varying the Higgs mass.

The result of the two-loop calculation in the considered 3-dimensional
models  suggests that the gap equation
approach is a reliable method to calculate the transverse
 propagator mass 
of the vector boson in the symmetric phase.
The two-loop calculation is a crucial test for the consistency of the whole method.
The physical interpretation and the connection to the masses of bound states
studied on the lattice requires further investigations \cite{noi}.
\bigskip
\bigskip
\bigskip

%\pagebreak 
\noindent
{\bf\Large Acknowledgments}

    \mbox{ }\\\noindent I would like to thank  W.~Buchm\"uller
and O.~Philipsen for highly instructive discussions and comments. I am also most grateful to O.~Tarasov for an independent calculation of eq.~(\ref{p2os})
as well as for helpful discussions.\\[2ex]

\newpage
\begin{appendix}

\section{Basic two-loop integrals\label{appa}}

In 3 Euclidean dimensions, the two-loop basic integrals are defined as,

\beqa
F(m_1,m_2,m_3,m_4,m_5) &=& \int\int {d^d k_1\over (2\pi )^d}{d^d k_2\over (2\pi )^d}
{1\over \left(k_1^2+m_1^2\right) \left(k_2^2+m_2^2\right)}\NO\\[1ex]
&&{1\over \left( (k_1-q)^2+m_3^2\right)\left( ( k_2-q)^2+m_4^2\right)\left( (k_1-k_2)^2+m_5^2\right)}\, ,\NO\\[1ex]
V(m_1,m_2,m_3,m_4) &=& \int\int {d^d k_1\over (2\pi )^d}{d^d k_2\over (2\pi )^d}
{1\over  \left(k_2^2+m_1^2\right)\left( (k_1-q)^2+m_2^2\right)}\NO\\[1ex]
&&{1\over \left( (k_2-q)^2+m_3^2\right)\left( (k_1-k_2)^2+m_4^2\right)}\, ,\NO\\[1ex]
I_{111}(q^2)(m_1,m_2,m_3) &=& \int\int {d^d k_1\over (2\pi )^d}{d^d k_2\over (2\pi )^d}
{1\over  \left(k_1^2+m_1^2\right)\left( (k_2-q)^2+m_2^2\right)}\, ,\NO\\[1ex]
&&{1\over \left( (k_1-k_2)^2+m_3^2\right)}\, ,\NO\\[1ex]
I_{211}(m_1,m_2,m_3) &=& -{\partial\over \partial m_1^2}
I_{111}(q^2)(m_1,m_2,m_3)\, ,\NO\\[1ex]
I_{121}(m_1,m_2,m_3) &=& -{\partial\over \partial m_2^2}
I_{111}(q^2)(m_1,m_2,m_3)\, ,\NO\\[1ex]
I_{112}(m_1,m_2,m_3) &=& -{\partial\over \partial m_3^2}
I_{111}(q^2)(m_1,m_2,m_3)\, ,\NO\\[1ex]
I_{111}(0)(m_1,m_2,m_3) &=& \int\int {d^d k_1\over (2\pi )^d}{d^d k_2\over (2\pi )^d}
{1\over  \left(k_1^2+m_1^2\right)\left( k_2^2+m_2^2\right)\left( (k_1-k_2)^2+m_3^2\right)}\label{besik}\, .
\eeqa
The one-loop integrals $A_0$ and $B_0$ are,
\beqa
A_0(m^2) &=& \int {d^3 k\over (2\pi)^3} {1\over k^2+m^2} = - {m\over 4\pi},\NO\\[1ex]
B_0(p^2,m_1^2,m_2^2) &=& \int {d^3 k\over (2\pi)^3} {1\over \left(k^2+m_1^2
\right)\left( (k+p)^2+m_2^2\right)} = {1\over 4\pi p} \mbox{arctan} {p\over m_1+m_2} \, .\label{aabbE}
\eeqa

Apart from the master integral $F(m_1,m_2,m_3,m_4,m_5)$, which has to be evaluated numerically, 
there exist analytic expressions for the basic integrals in $d=3-2\e$ dimensions
\cite{raj}.
%The results are summarized in appendix \ref{ap3}.
For $F$ a one-dimensional integral remains.
%In appendix \ref{ap5} we evaluate $F(m_1,m_2,m_3,m_4,m_5)$ numerically for
%some special mass cases.

%\input{appbp}

%\input{appdp}
\section{Two-loop results in the non-linear $\s$-model\label{ap2}}

In section 2, the transverse two-loop vector boson
self-energy is calculated in $d=3-2\e $ dimensions on as well
 as off mass-shell in the non-linear $\s$-model.
Here the on-shell result of the reduction is given in arbitrary dimension
$d$.
The basic integrals  are defined in eq.~(\ref{besik}). 

\beqa
&&{1\over g^4} \Pi^{2-loop}_T(p^2) = \NO\\[1ex]
&&{3\over 16} m^4 {176 d - 245\over d-1} F(m,m,m,m,m)\NO\\[1ex]
&&-{3\over 16}m^2 {144 d^3- 712 d^2 + 1241 d -760\over (d-1)^2} V(m,m,m,m)\NO\\[1ex]
&&-{1\over 48}{10800 d^4 - 70632 d^3 +165227 d^2 - 166654 d + 61752\over (d-1)^2 (3d-4)} I(p^2=-m^2)(m,m,m)
\NO\\[1ex]
&&-{3\over 16}{(d-2) (32 d^3 - 312 d^2 + 656 d -405)\over (d-1)^2}  I(0)(m,m,m)
\NO\\[1ex]
&&+{3\over 32}{32 d^2 - 148 d + 155\over (d-1)^2} B(p^2=-m^2,m^2,m^2) B(p^2=-m^2,m^2,m^2)
\NO\\[1ex]
&&-{3\over 4} {16 d^4 - 188 d^3 + 668 d^2 - 940 d +465\over (d-1)^2} B(p^2=-m^2,m^2,m^2) A(m^2)
\NO\\[1ex]
&&-{1\over 8 m^2} {(2d-3) (24 d^5 -164 d^4 +452 d^3 - 680 d^2 +597 d -242)\over (d-1)^2 (3d-4) }A(m^2) A(m^2)\, ,
\eeqa

Switching back to $d=3-2\e $, we write down the result for the off-shell
transverse two-loop self-energy of the vector field in Feynman gauge. It is

\beqa
&&{1\over g^4} \Pi^{2-loop}_T(p^2) = \NO\\[1ex]
&&\left( {257\over 16} m^4 - {351\over 32} p^2 m^2 - {1\over 2} p^4 \right) F(m,m,m,m,m)\NO\\[1ex]
&& + \left( - {259\over 8} m^2 - {1265\over 64} p^2  - {261\over 32} {p^4\over m^2} \right) V(m,m,m,m)\NO\\[1ex]
&& + \left( {8163\over 20} m^8 - {4607\over 80} p^2 m^6  - {12183\over 20} p^4 m^4 - {12243\over 80} p^6 m^2 - {77\over 8} p^8 \right){I_{211}(m,m,m)\over  m^2 p^2 (p^2+4 m^2)}\NO\\[1ex]
&& + \left( - {279\over 4} m^6 + {1409\over 48} p^2 m^4  + {53279\over 960} p^4 m^2 + {3923\over 480} p^6\right.\NO\\[1ex]
&&\left.+ {8717\over 20} m^6 \e - {14647\over 60} p^2 m^4 \e - {225067\over 600} p^4 m^2 \e - {5473\over 100} p^6 \e\right)
 {I_{111}(p^2)(m,m,m)\over  m^2 p^2 (p^2+4 m^2)} \NO\\[1ex]
&& + \left( + {279\over 4} m^6 - {507\over 16} p^2 m^4  - {3585\over 64} p^4 m^2 - {261\over 32} p^6\right.\NO\\[1ex]
&&\left.- {655\over 4} m^6 \e - {51\over 8} p^2 m^4 \e  + {537\over 8} p^4 m^2 \e + {35\over 4} p^6\e \right)
 {I_{111}(0)(m,m,m)\over  m^2 p^2 (p^2+4 m^2)} \NO\\[1ex]
&&+ \, \mbox{products of one-loop integrals}\, .
\eeqa
The  coefficients in front
of the generic basic two-loop integrals coincide in unitary and
Feynman gauge on mass-shell, see 
eq.~(\ref{p2os}).
More detailed results for single two-loop diagrams can be found in 
\cite{frdiss}.
\def\sourco#1{\;\parbox{15mm}{\setlength{\unitlength}{0.15mm}
\begin{picture}(70,50)\thicklines#1
\end{picture}}}
\def\dreio{\sourco{\put(0,25){\line(1,0){25}}\put(50,25){\circle{50}}\put(50,0){\line(0,1){50}}\put(75,25){\line(1,0){25}}}}

\section{Two-loop results in the $SU(2)$ Higgs model\label{ap4}}

\begin{figure}
\begin{center}
\psfig{file=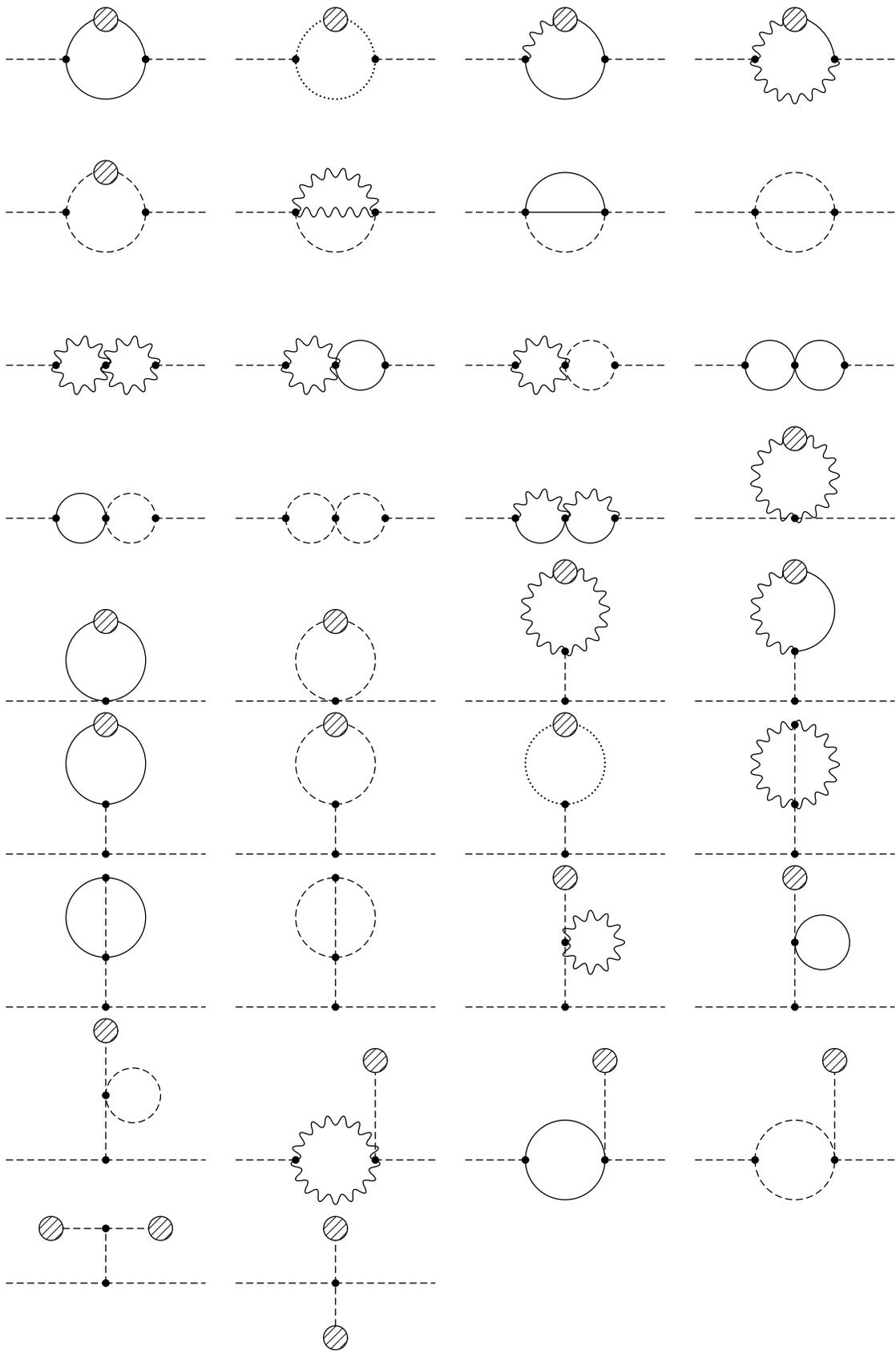}
\caption{ \it Two-loop diagrams for the Higgs self-energy\label{fiy}}
\end{center}
\end{figure}

\begin{figure}
\begin{center}
\psfig{file=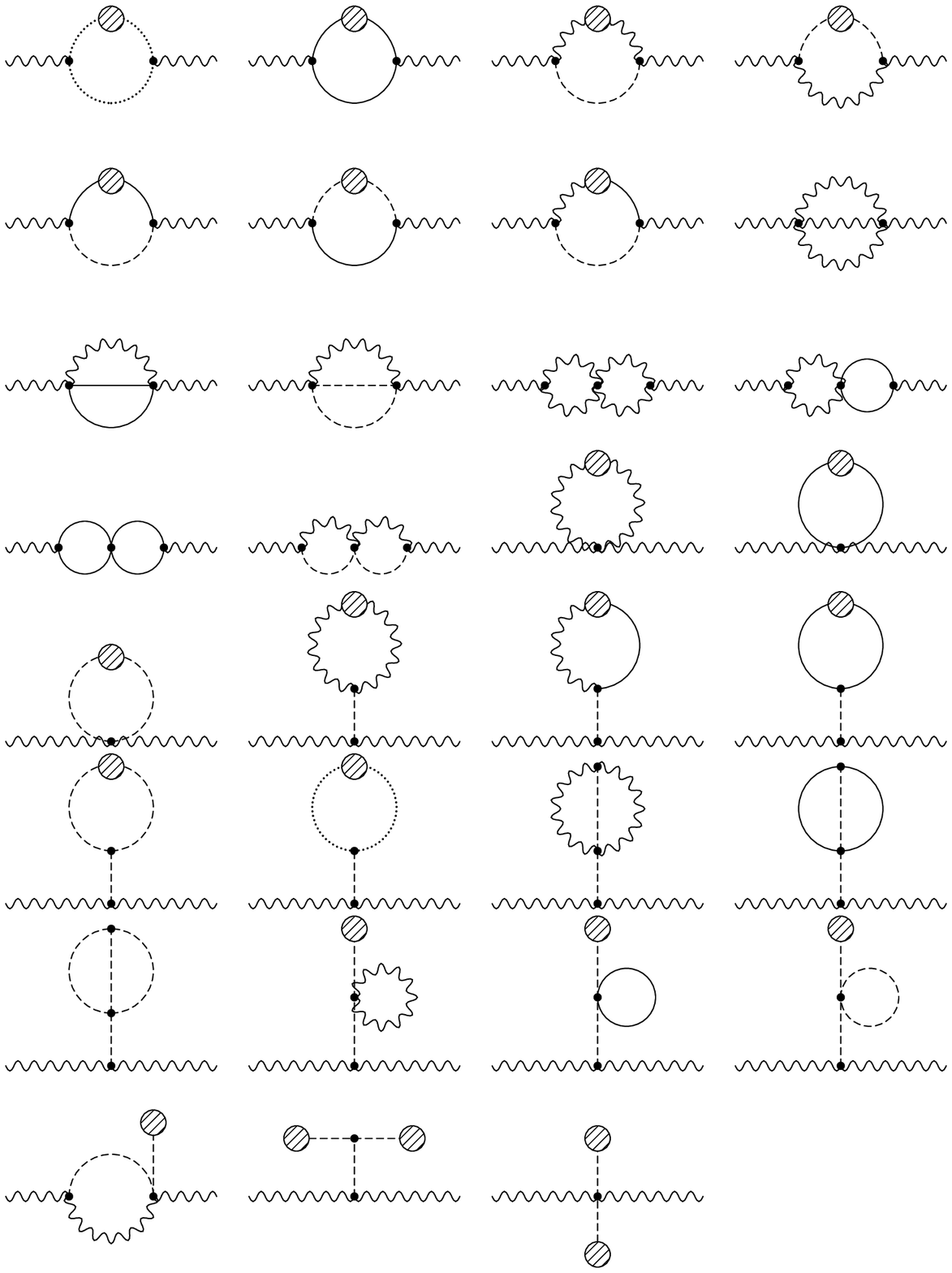}
\caption{ \it Two-loop diagrams for the vector boson self-energy\label{fix}}
\end{center}
\end{figure}

\begin{figure}
\begin{center}
\psfig{file=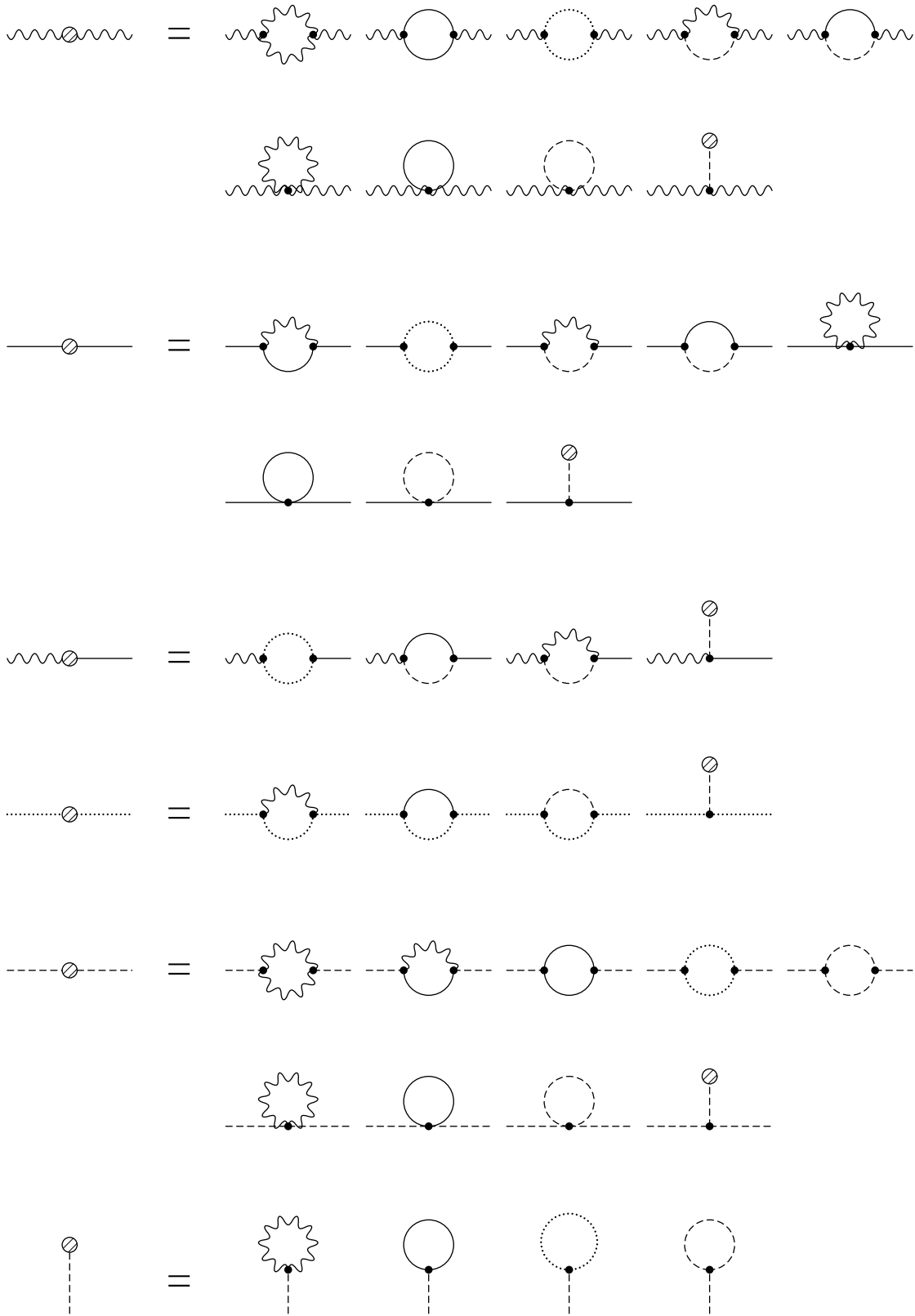}
\caption{ \it One-loop self-energy insertions into two-loop diagrams\label{fiz}}
\end{center}
\end{figure}

The on-shell value of the  Higgs boson self-energy  and of the transverse vector boson self-energy to two loops is given in unitary and Feynman gauge 
in $3-2\e $ dimensions,
neglecting resummation counter-terms.
The coefficients in front of the generic basic two-loop integrals are identical
 in both gauges.
In the unitary gauge result, we therefore write only the part containing products of one-loop integrals.
With the following formulae and the one-loop
results in \cite{BP} and in sect.~\ref{21},
the gauge-invariance of the pole of the Higgs and the vector propagator
can  easily be proved to two loops.
The relevant diagrams are given in figs.~\ref{fiy}, \ref{fix}  and \ref{fiz}.

The two-loop self-energy for the Higgs field in Feynman gauge reads,

\beqa
&&\Sigma^{\xi = 1}(p^2=-M^2)=\left( - {189\over 4} m^4 +{81\over 4} M^2 m^2-{33\over 16} M^4+{3\over 8} {M^6\over m^2}\right) F(m,m,m,m,m)\NO\\[1ex]
&&+\left( 3 m^4 -3 M^2 m^2-{3\over 4} M^4+{3\over 4} {M^6\over m^2}+{3\over 32} {M^8\over m^4}\right) F(m,m,m,m,M)\NO\\[1ex]
&&+\left( 9 M^2 m^2-{27\over 4} M^4+{9\over 16} {M^8\over m^4}\right) F(m,M,m,M,m)\NO\\[1ex]
&&+{81\over 32} {M^8\over m^4} F(M,M,M,M,M)\NO\\[1ex]
&&+\left(  {189\over 8} m^2 +6 M^2-{213\over 64} {M^4\over m^2}\right) V(m,m,m,m)\NO\\[1ex]
&&+\left(  -3 m^2 +{3\over 2} M^2+3 {M^4\over m^2}-{3\over 2} {M^6\over m^4}+{15\over 64} {M^8\over m^6}\right) V(m,m,m,M)\NO\\[1ex]
&&+\left(  -{9\over 8} M^2+{45\over 16} {M^4\over m^2}-{9\over 64} {M^6\over m^4}\right) V(M,m,M,m)\NO\\[1ex]
&&-{27\over 64} {M^6\over m^4} V(M,M,M,M)\NO\\[1ex]
&&+\left( -{567\over 2} m^8 + {2097\over 4} M^2 m^6 -{5061\over 16} M^4 m^4+{651\over 8} M^6 m^2 -{93\over 16} M^8\right) {I_{211}(m,m,m)\over M^2 m^2 (M^2-4m^2)}\NO\\[1ex]
&&+\left( 24 m^{10}- 18 M^2 m^8 -{57\over 4} M^6 m^4+{81\over 8} M^8 m^2 -{15\over 8} M^{10}\right) {I_{211}(M,m,m)\over M^2 m^4 (M^2-4m^2)}\NO\\[1ex]
&&+\left( -{189\over 4} m^6 + {459\over 8} M^2 m^4 -{111\over 4} M^4 m^2+{309\over 64} M^6\right.\NO\\[1ex]
&&\left. +{675\over 2} m^6\e - {1449\over 4} M^2 m^4\e +{2727\over 16} M^4 m^2\e-{1953\over 64} M^6\e\right) {I_{111}(p^2=-M^2)(m,m,m)\over M^2 m^2 (M^2-4m^2)}\NO\\[1ex]
&&+\left( 6 m^{10} - {15\over 2} M^2 m^8 +{27\over 8} M^4 m^6-{27\over 8} M^6 m^4 +{105\over 64} M^8 m^2\right.\NO\\[1ex]
&&\left. -{15\over 64} M^{10} - 42 m^{10}\e + {45\over 2} M^2 m^8\e -{87\over 8} M^4 m^6\e+{45\over 4} M^6 m^4\e\right.\NO\\[1ex]
&&\left. -{171\over 32} M^8 m^2\e+{51\over 64} M^{10}\e\right) {I_{111}(p^2=-M^2)(M,m,m)\over M^2 m^6 (M^2-4m^2)}\NO\\[1ex]
&&+\left( - {21\over 64} {M^4\over  m^4}+ {81\over 32} {M^4\over  m^4}\e\right) I_{111}(p^2=-M^2)(M,M,M)\NO\\[1ex]
&&+\left( {189\over 4} m^6 - {711\over 8} M^2 m^4 +{261\over 8} M^4 m^2-{261\over 64} M^6\right.\NO\\[1ex]
&&\left. -{297\over 2} m^6\e + {909\over 4} M^2 m^4\e -{837\over 16} M^4 m^2\e+{225\over 64} M^6\e\right) {I_{111}(0)(m,m,m)\over M^2 m^2 (M^2-4m^2)}\NO\\[1ex]
&&+\left( -6 m^{10} + {27\over 2} M^2 m^8 -{87\over 8} M^4 m^6+{45\over 8} M^6 m^4 -{117\over 64} M^8 m^2\right.\NO\\[1ex]
&&\left. +{15\over 64} M^{10} + 18 m^{10}\e - {63\over 2} M^2 m^8\e +{111\over 8} M^4 m^6\e-{9\over 8} M^6 m^4\e \right.\NO\\[1ex]
&&\left. -{27\over 32} M^8 m^2\e+{9\over 64} M^{10}\e\right) {I_{111}(0)(M,m,m)\over M^2 m^6 (M^2-4m^2)}\NO\\[1ex]
&&+\left(  {9\over 64} {M^4\over  m^4}- {9\over 32} {M^4\over  m^4}\e\right) I_{111}(0)(M,M,M)\NO\\[1ex]
&&+\left( - {3\over 2} m^2 - {9\over 16} M^2 - {9\over 16} {M^4\over m^2} - {15\over 64} {M^6\over m^4}\right) B(p^2=-M^2)(m^2,m^2) B(p^2=-M^2)(m^2,m^2)\NO\\[1ex]
&&+\left( - {9\over 4} M^2 + {9\over 16} {M^4\over m^2} - {9\over 32} {M^6\over m^4}\right) B(p^2=-M^2)(m^2,m^2) B(p^2=-M^2)(M^2,M^2)\NO\\[1ex]
&&- {27\over 64} {M^6\over m^4} B(p^2=-M^2)(M^2,M^2) B(p^2=-M^2)(M^2,M^2)\NO\\[1ex]
&&+\left( 36 m^{10} + 60 M^2 m^8 - {99\over 2} M^4 m^6 + {75\over 4} M^6 m^4
\right.\NO\\[1ex]
&&\left. - {57\over 32} M^8 m^2 - {15\over 64} M^{10}\right)
 {B(p^2=-M^2)(m^2,m^2) A(m^2)\over M^2 m^6 (M^2-4 m^2)}\NO\\[1ex]
&&+\left( - 6 m^{10} + 18 M^2 m^8 - {39\over 4} M^4 m^6 + {45\over 8} M^6 m^4
\right.\NO\\[1ex]
&&\left. - {63\over 32} M^8 m^2 + {15\over 64} M^{10}\right)
 {B(p^2=-M^2)(m^2,m^2) A(M^2)\over M^2 m^6  (M^2-4 m^2)}\NO\\[1ex]
&&+\left( {27\over 4} M^2 m^4 + {27\over 8} M^4 m^2 - {27\over 16} M^6\right)
 {B(p^2=-M^2)(M^2,M^2) A(m^2)\over  m^4  (M^2-4 m^2)}\NO\\[1ex]
&&+\left( 27 m^8 - {39\over 4} M^2 m^6 + {33\over 32} M^6 m^2 - {15\over 64} M^8\right){A(m^2) A(m^2)\over M^2 m^6 (M^2-4 m^2)}\NO\\[1ex]
&&+\left( {57\over 4} m^6 - {51\over 32} M^4 m^2 + {15\over 64} M^6\right){A(m^2) A(M^2)\over  m^6 (M^2-4 m^2)}\NO\\[1ex]
&&- {9\over 16} {M^2\over m^4} A(M^2) A(M^2)\label{hiegs}\, .
\eeqa

\bigskip
\noindent The vector boson self-energy in Feynman gauge is to two-loop order,

\beqa
&&\Pi_T^{\xi = 1}(p^2=-m^2)= {849\over 32} m^4  F(m,m,m,m,m)\NO\\[1ex]
&&+\left( - {63\over 8} m^4 + {27\over 8} M^2 m^2 - {11\over 32} M^4 + {1\over 16} {M^6\over m^2}\right) F(m,m,m,m,M)\NO\\[1ex]
&&+\left( - {63\over 2} m^4 + {27\over 2} M^2 m^2 - {11\over 8} M^4 + {1\over 4} {M^6\over m^2}\right) F(M,m,m,m,m)\NO\\[1ex]
&&+\left(  m^4 - M^2 m^2 - {1\over 4} M^4 + {1\over 4} {M^6\over m^2}+ {1\over 32} {M^8\over m^4}\right) F(M,m,m,M,m)\NO\\[1ex]
&&+\left(  {3\over 2} M^2 m^2 - {9\over 8} M^4 + {3\over 32} {M^8\over m^4}\right) F(m,m,M,M,M)\NO\\[1ex]
&&+\left(  - {2115\over 64} m^2 + M^2 - {1\over 4} {M^4\over m^2}\right) V(m,m,m,m)\NO\\[1ex]
&&+\left(   {63\over 8} m^2 + 2 M^2 - {71\over 64} {M^4\over m^2}\right) V(m,M,m,m)\NO\\[1ex]
&&+\left( {3\over 2} {m^4\over M^2} + {1\over 2} m^2 - {9\over 8} M^2 - {3\over 16} {M^4\over m^2} - {17\over 128} {M^6\over m^4}\right) V(m,m,M,m)\NO\\[1ex]
&&+\left(  {63\over 16} m^2 + M^2 - {71\over 128} {M^4\over m^2}\right) V(M,m,m,m)\NO\\[1ex]
&&+\left( - {1\over 2} m^2 + {1\over 4} M^2 + {1\over 2} {M^4\over m^2} - {1\over 4} {M^6\over m^4} + {5\over 128} {M^8\over m^6}\right) V(M,M,m,m)\NO\\[1ex]
&&+\left( {3\over 8} M^2 + {3\over 4} {M^4\over m^2} - {15\over 64} {M^6\over m^4}\right) V(m,M,M,M)\NO\\[1ex]
&&+\left( {189\over 4} m^8 - {699\over 8} M^2 m^6 + {1687\over 32} M^4 m^4 - {217\over 16} M^6 m^2 + {31\over 32} M^8\right) {I_{211}(m,m,M)\over  m^4 (M^2-4m^2)}\NO\\[1ex]
&&+\left( -4 m^{10} + {9\over 2} M^4 m^6 + {1\over 2} M^6 m^4 - {21\over 16} M^8 m^2 + {5\over 16} M^{10}\right) {I_{211}(M,m,M)\over  m^6 (M^2-4m^2)}\NO\\[1ex]
&&+\left( {3\over 2} m^8 + {1021\over 16} M^2 m^6 - {997\over 64} M^4 m^4 - {1\over 2} M^6 m^2 + {13\over 128} M^8\right.\NO\\[1ex]
&&\left. - {21\over 2} m^8\e - {1881\over 4} M^2 m^6\e + {1851\over 16} M^4 m^4\e + {5\over 8} M^6 m^2\e - {5\over 128} M^8\e\right) {I_{111}(p^2=-m^2)(m,m,m)\over M^2 m^4 (M^2-4m^2)}\NO\\[1ex]
&&+\left( - 15 m^6 + {799\over 64} M^2 m^4 - {559\over 128} M^4 m^2 + {31\over 64} M^6\right.\NO\\[1ex]
&&\left. + {1617\over 16} m^6\e - {2251\over 32} M^2 m^4\e + {737\over 32} M^4 m^2\e - {41\over 16} M^6 \e\right) {I_{111}(p^2=-m^2)(M,m,m)\over  m^4 (M^2-4m^2)}\NO\\[1ex]
&&+\left( - {1\over 2} m^8 + {9\over 8} M^2 m^6 - {13\over 16} M^4 m^4 + {1\over 16} M^6 m^2 + {5\over 128} M^8\right.\NO\\[1ex]
&&\left. -  2 m^8\e - {23\over 4} M^2 m^6\e + {9\over 8} M^4 m^4\e + {1\over 64} M^6 m^2\e - {3\over 32} M^8\e\right) {I_{111}(p^2=-m^2)(M,M,m)\over  m^6 (M^2-4m^2)}\NO\\[1ex]
&&+\left( - {63\over 2} m^6 - {951\over 16} M^2 m^4 +{483\over 32} M^4 m^2 + {87\over 128} M^6\right.\NO\\[1ex]
&&\left.  99 m^6\e + {1647\over 16} M^2 m^4\e - {1101\over 32} M^4 m^2\e + {3\over 32} M^6\e\right) {I_{111}(0)(m,m,m)\over M^2 m^2 (M^2-4m^2)}\NO\\[1ex]
&&+\left(  {9\over 2} m^6 + {17\over 2} M^2 m^4 - 9 M^4 m^2 + {7\over 2} M^6
- {41\over 128} {M^8\over m^2} - {5\over 128} {M^{10}\over m^4}
\right.\NO\\[1ex]
&&\left. - {27\over 2} m^6\e - {29\over 2} M^2 m^4\e + {49\over 4} M^4 m^2\e - {21\over 4} M^6\e
 + {149\over 128} {M^8\over m^2}\e - {1\over 16} {M^{10}\over m^4}\e
\right) {I_{111}(0)(M,m,m)\over M^2 m^2 (M^2-4m^2)}\NO\\[1ex]
&&+\left(  {9\over 8} M^2 m^4 + {3\over 4} M^4 m^2 - {21\over 64} M^6
 - {3\over 4} M^2 m^4\e + {3\over 16} M^4 m^2\e + {9\over 64} M^6\e
 \right) {I_{111}(0)(M,M,M)\over  m^4 (M^2-4m^2)}\NO\\[1ex]
&&+\left(  {45\over 16} m^2 + {1\over 4} M^2 - {1\over 16} {M^4\over m^2}\right) B(p^2=-m^2)(m^2,m^2) B(p^2=-m^2)(m^2,m^2)\NO\\[1ex]
&&+\left(  {9\over 2} m^2 - {5\over 4} M^2 + {1\over 2} {M^4\over m^2}\right) B(p^2=-m^2)(m^2,m^2) B(p^2=-m^2)(M^2,m^2)\NO\\[1ex]
&&+\left( - {1\over 2} m^2 + {5\over 8} M^2 - {1\over 8} {M^4\over m^2}\right) B(p^2=-m^2)(M^2,m^2) B(p^2=-m^2)(M^2,m^2)\NO\\[1ex]
&&+\left( - 54  m^6 - {963\over 8} M^2 m^4 + {63\over 2} M^4 m^2 + {63\over 64} M^6 \right)
 {B(p^2=-m^2)(m^2,m^2) A(m^2)\over M^2 m^2 (M^2-4 m^2)}\NO\\[1ex]
&&+\left(  {63\over 4}  m^6 - {207\over 8} M^2 m^4 + {135\over 16} M^4 m^2 - {63\over 64} M^6 \right)
 {B(p^2=-m^2)(m^2,m^2) A(M^2)\over  M^2 m^2 (M^2-4 m^2)}\NO\\[1ex]
&&+\left(  9 m^{10} + {9\over 2} M^2 m^8 - {35\over 4} M^4 m^6 + {9\over 2} M^6 m^4
\right.\NO\\[1ex]
&&\left. - {35\over 64} M^8 m^2 - {5\over 128} M^{10}\right)
 {B(p^2=-m^2)(M^2,m^2) A(m^2)\over M^2 m^6  (M^2-4 m^2)}\NO\\[1ex]
&&+\left( - m^{10} + M^2 m^8 - {9\over 8} M^4 m^6 + {7\over 8} M^6 m^4
\right.\NO\\[1ex]
&&\left. - {5\over 16} M^8 m^2 + {5\over 128} M^{10}\right)
 {B(p^2=-m^2)(M^2,m^2) A(M^2)\over M^2 m^6  (M^2-4 m^2)}\NO\\[1ex]
&&+\left(  18 m^{10} - 39 M^2 m^8 + {51\over 4} M^4 m^6 + {11\over 32} M^6 m^4
\right.\NO\\[1ex]
&&\left. - {11\over 32} M^8 m^2 + {5\over 128} M^{10}\right)
{ A(m^2) A(m^2)\over M^4 m^6  (M^2-4 m^2)}\NO\\[1ex]
&&+\left(  3 m^{10} + M^2 m^8 - {19\over 8} M^4 m^6 - {17\over 16} M^6 m^4
\right.\NO\\[1ex]
&&\left. + {29\over 64} M^8 m^2 - {5\over 128} M^{10}\right)
 {A(m^2) A(M^2)\over M^4 m^6  (M^2-4 m^2)}\NO\\[1ex]
&&+\left( {1\over 2} m^6 + {1\over 8} M^2 m^4 + {11\over 32} M^4 m^2 - {7\over 64} M^6\right){A(M^2) A(M^2)\over M^2 m^4 (M^2-4 m^2)}\label{veict}\, .
\eeqa

\bigskip
\noindent In unitary gauge, the products of one-loop integrals in the two-loop 
self-energy of the Higgs field read,

\beqa
&&\Sigma^{\xi = \infty}(p^2=-M^2) = \, .\, .\, .\NO\\[1ex]
&&+\left( - {3\over 2} m^2 + {27\over 16} M^2 - {27\over 16} {M^4\over m^2} + {3\over 64} {M^6\over m^4}\right) B(p^2=-M^2)(m^2,m^2) B(p^2=-M^2)(m^2,m^2)\NO\\[1ex]
&&+\left( - {9\over 4} M^2 + {9\over 16} {M^4\over m^2} + {9\over 16} {M^6\over m^4}\right) B(p^2=-M^2)(m^2,m^2) B(p^2=-M^2)(M^2,M^2)\NO\\[1ex]
&&- {27\over 64} {M^6\over m^4} B(p^2=-M^2)(M^2,M^2) B(p^2=-M^2)(M^2,M^2)\NO\\[1ex]
&&+\left( 36 m^{10} + 51 M^2 m^8 - {207\over 4} M^4 m^6 + {33\over 2} M^6 m^4
\right.\NO\\[1ex]
&&\left. - {15\over 16} M^8 m^2 - {15\over 64} M^{10}\right)
 {B(p^2=-M^2)(m^2,m^2) A(m^2)\over M^2 m^6 (M^2-4 m^2)}\NO\\[1ex]
&&+\left( - 6 m^{10} + 18 M^2 m^8 - {39\over 4} M^4 m^6 + {27\over 8} M^6 m^4
\right.\NO\\[1ex]
&&\left. - {45\over 32} M^8 m^2 + {15\over 64} M^{10}\right)
 {B(p^2=-M^2)(m^2,m^2) A(M^2)\over M^2 m^6  (M^2-4 m^2)}\NO\\[1ex]
&&+\left( {27\over 4} M^2 m^4 - {27\over 32} M^6\right)
 {B(p^2=-M^2)(M^2,M^2) A(m^2)\over  m^4  (M^2-4 m^2)}\NO\\[1ex]
&&+\left( 27 m^8 - {75\over 4} M^2 m^6 + {51\over 32} M^6 m^2 - {15\over 64} M^8\right){A(m^2) A(m^2)\over M^2 m^6 (M^2-4 m^2)}\NO\\[1ex]
&&+\left( {57\over 4} m^6 - {9\over 4} M^2 m^4 - {33\over 32} M^4 m^2 + {15\over 64} M^6\right){A(m^2) A(M^2)\over  m^6 (M^2-4 m^2)}\NO\\[1ex]
&&- {9\over 16} {M^2\over m^4} A(M^2) A(M^2)\, ,
\eeqa

\bigskip
\noindent
and for the vector field,

\beqa
&&\Pi_T^{\xi = \infty}(p^2=-m^2) = \, .\, .\, .\NO\\[1ex]
&&+\left( - {99\over 32} m^2 + {1\over 4} M^2 - {1\over 16} {M^4\over m^2}\right) B(p^2=-m^2)(m^2,m^2) B(p^2=-m^2)(m^2,m^2)\NO\\[1ex]
&&+\left(  {21\over 4} m^2 - {13\over 8} M^2 + {19\over 32} {M^4\over m^2}\right) B(p^2=-m^2)(m^2,m^2) B(p^2=-m^2)(M^2,m^2)\NO\\[1ex]
&&+\left( - {1\over 2} m^2 + {5\over 8} M^2 - {1\over 8} {M^4\over m^2}\right) B(p^2=-m^2)(M^2,m^2) B(p^2=-m^2)(M^2,m^2)\NO\\[1ex]
&&+\left( - 63  m^6 - {555\over 8} M^2 m^4 + {315\over 16} M^4 m^2 + {57\over 64} M^6 \right)
 {B(p^2=-m^2)(m^2,m^2) A(m^2)\over M^2 m^2 (M^2-4 m^2)}\NO\\[1ex]
&&+\left(  {63\over 4}  m^6 - {219\over 8} M^2 m^4 + {135\over 16} M^4 m^2 - {57\over 64} M^6 \right)
 {B(p^2=-m^2)(m^2,m^2) A(M^2)\over  M^2 m^2 (M^2-4 m^2)}\NO\\[1ex]
&&+\left(  9 m^{10} - {3\over 2} M^2 m^8 - {17\over 4} M^4 m^6 + 3 M^6 m^4
\right.\NO\\[1ex]
&&\left. - {23\over 64} M^8 m^2 - {5\over 128} M^{10}\right)
 {B(p^2=-m^2)(M^2,m^2) A(m^2)\over M^2 m^6  (M^2-4 m^2)}\NO\\[1ex]
&&+\left( - m^{10} + M^2 m^8 - {9\over 8} M^4 m^6 + {7\over 8} M^6 m^4
\right.\NO\\[1ex]
&&\left. - {5\over 16} M^8 m^2 + {5\over 128} M^{10}\right)
 {B(p^2=-m^2)(M^2,m^2) A(M^2)\over M^2 m^6  (M^2-4 m^2)}\NO\\[1ex]
&&+\left(  18 m^{10} - 57 M^2 m^8 + {81\over 4} M^4 m^6 + {11\over 32} M^6 m^4
\right.\NO\\[1ex]
&&\left. - {17\over 32} M^8 m^2 + {5\over 128} M^{10}\right)
{ A(m^2) A(m^2)\over M^4 m^6  (M^2-4 m^2)}\NO\\[1ex]
&&+\left(  3 m^{10} + M^2 m^8 - {43\over 8} M^4 m^6 - {17\over 16} M^6 m^4
\right.\NO\\[1ex]
&&\left. + {41\over 64} M^8 m^2 - {5\over 128} M^{10}\right)
 {A(m^2) A(M^2)\over M^4 m^6  (M^2-4 m^2)}\NO\\[1ex]
&&+\left( {1\over 2} m^6 + {1\over 8} M^2 m^4 + {11\over 32} M^4 m^2 - {7\over 64} M^6\right){A(M^2) A(M^2)\over M^2 m^4 (M^2-4 m^2)}\, .
\eeqa

\begin{figure}
\begin{center}
\psfig{file=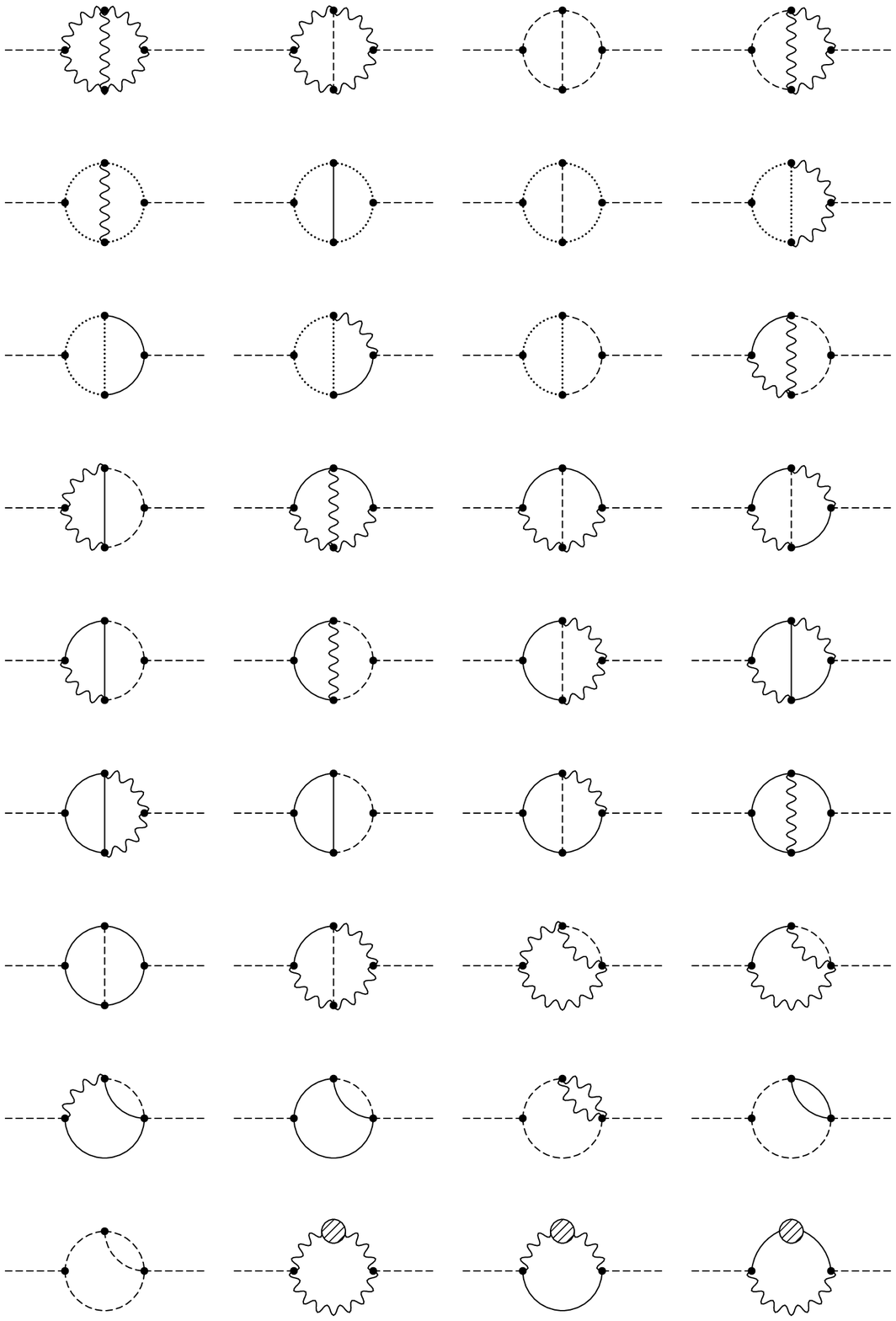}
\end{center}
\end{figure}

\input{psfig}

\input{psfig}
\begin{figure}
\begin{center}
\psfig{file=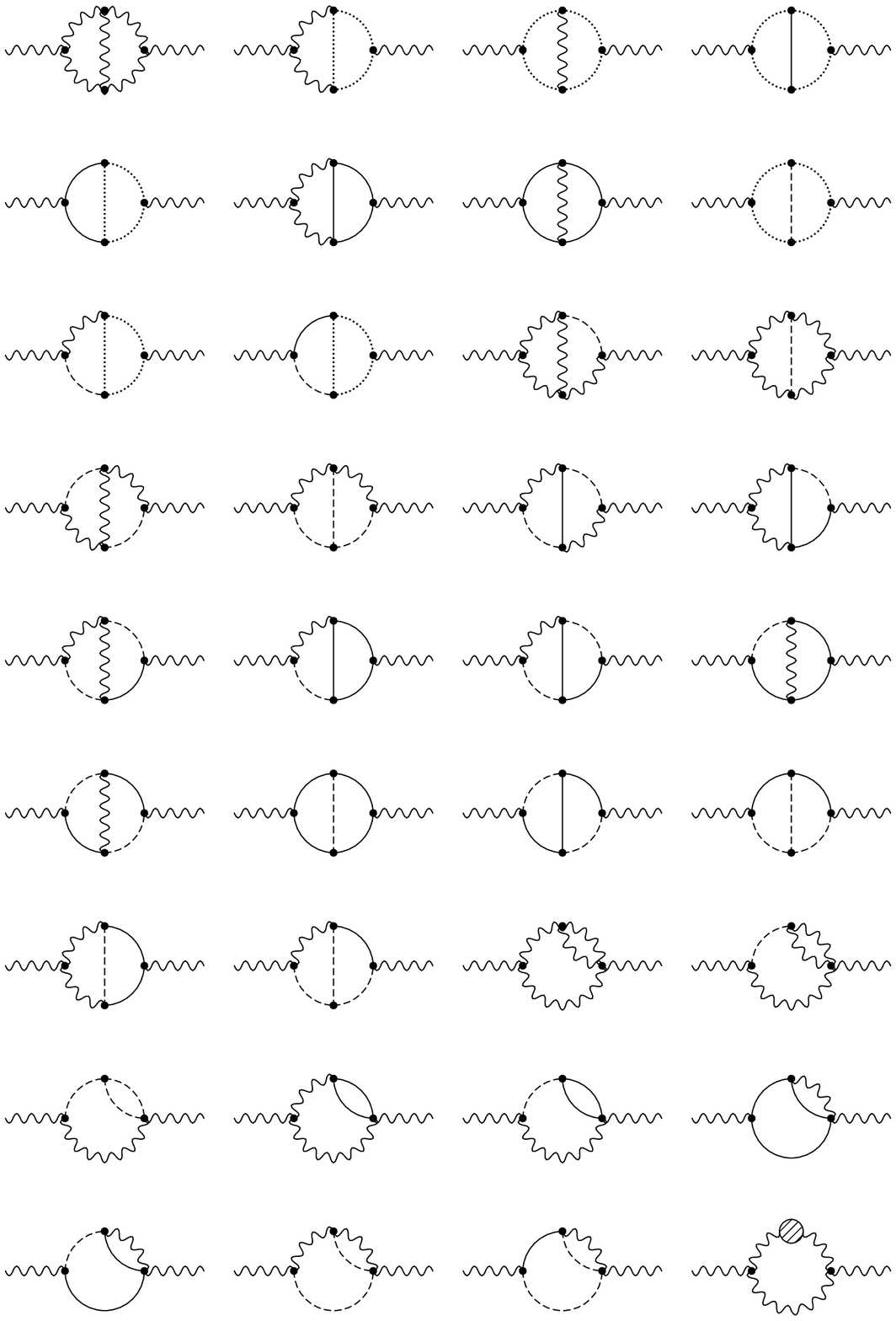}

\end{center}
\end{figure}

\input{psfig}

\input{psfig}

\end{appendix}


\newpage
 \begin{thebibliography}{99}
 % \input{bibp.tex}
 
\bibitem{halp} 
  B.~I.~Halperin, T.~C.~Lubensky and S.-K.~Ma, \pl{32}{74}{292}




\bibitem{7-9o}
  D.~A.~Kirzhnits and A.~D.~Linde, \pl{72}{72}{471}
 % S.~Weinberg, \pr{9}{74}{3357}\\
  

\bibitem{vier} 
 
 A.~D.~Linde, \pl{96}{80}{289};\\
 D.~Gross, R.~Pisarski and L.~Yaffe, Rev.~Mod.~Phys.~53 (1981) 43 
\bibitem{fein} 
  R.~P.~Feynman, \np{188}{81}{479}

\bibitem{funf} 
K.~Kajantie, M.~Laine, K.~Rummukainen and M.~Shaposhnikov, \np{466}{96}{189}



\bibitem{ptw} 
  O.~Philipsen, M.~Teper and H.~Wittig, \np{469}{96}{445}

\bibitem{kar} 
  F.~Karsch, T.~Neuhaus, A.~Patk\'{o}s and J.~Rank, \np{474}{96}{217}

%\bibitem{brps} 
%  W.~Buchm\"uller and O.~Philipsen, \pl{397}{97}{112}

%\bibitem{bs} 
%  B.~Grossmann, S.~Gupta, U.~M.~Heller and F.~Karsch, \np{417}{94}{289}\\
%  H.-G.~Dosch, J.~Kripfganz, A.~Laser and M.~G.~Schmidt, \pl{365}{95}{213}

%\bibitem{laph} 
 % M.~Laine and O.~Philipsen, {\tt hep-lat/9711022}
\bibitem{BP} 
  W.~Buchm\"uller and O.~Philipsen, \np{443}{95}{47}

\bibitem{frenk}
  F.~Eberlein, \pl{439}{98}{130}

%\bibitem{tep} 
%  M.~Teper, \pl{289}{92}{115}


%\bibitem{rebhan} 
%  A.~K.~Rebhan, \pr{48}{93}{3967}

%\bibitem{mont} 
%  I.~Montvay and G.~M\"unster,
%  {\it{Quantum Fields on a Lattice}}, (Cambridge University Press, 1994)

\bibitem{JP1}
  R.~Jackiw and S.-Y.~Pi, \pl{368}{96}{131}, \pl{403}{97}{297}
%\bibitem{jak} 
%  L.~Dolan and R.~Jackiw, \pr{9}{74}{2904}

%\bibitem{oweh} 
%  O.~Philipsen, Doctoral Thesis

%\bibitem{ryd} 
%  L.~H.~Ryder, 
% {\it{Quantum Field Theory}}, (Cambridge University Press, 1985)

%\bibitem{5678} 
%  B.~W.~Lee and J.~Zinn-Justin, \pr{5}{72}{3121}, 3137, 3155, {\bf D 7} (1973)% 1049\\
%  K.~Fujukawa, B.~W.~Lee and A.~I.~Sanda, \pr{6}{72}{2923}\\
%  E.~S.~Abers and B.~W.~Lee, Phys.~Rep.~{\bf 9} (1973) 1\\
%  S.~Weinberg, \pr{7}{73}{2887}\\
%  T.~D.~Lee and C.~N.~Yang, Phys.~Rev.~{\bf 128} (1962) 885

%\bibitem{zero} 
%  I.~Gerstein, R.~Jackiw, B.~Lee and S.~Weinberg, \pr{3}{71}{2486}

%\bibitem{AN} 
%  G.~Alexanian and V.~P.~Nair, \pl{352}{95}{435}




%\bibitem{C}
%  J.~M.~Cornwall, \pr{57}{98}{3694}

\bibitem{Tar}
  O.~V.~Tarasov, \np{502}{97}{455}, and refernces therein


\bibitem{verm}
  J.~A.~M.~Vermaseren, {\it Symbolic manipulation with FORM.} Amsterdam, Computer Algebra Nederland, 1991
\bibitem{jak} 
  L.~Dolan and R.~Jackiw, \pr{9}{74}{2904}

\bibitem{rebhan} 
  A.~K.~Rebhan, \pr{48}{93}{3967}
\bibitem{AN} 
  G.~Alexanian and V.~P.~Nair, \pl{352}{95}{435}

\bibitem{C}
  J.~M.~Cornwall, \pr{57}{98}{3694}

\bibitem{tepi}
  M.~Teper, \pl{311}{93}{223}\\
  D.~Karabali, C.~Kim and V.~P.~Nair, \pl{434}{98}{103}
%\bibitem{tep} 
%  M.~Teper, \pl{289}{92}{115}

%\bibitem{ote}
%  O.~Philipsen, M.~Teper and H.~Wittig, \np{469}{96}{445}

\bibitem{raj}
  A.~K.~Rajantie, \np{480}{96}{729}, {\bf B 513} (1998) 761
%\bibitem{JP2}
%  R.~Jackiw and S.-Y.~Pi, \pl{403}{97}{297}



\bibitem{frdiss}
  F.~Eberlein, DESY-THESIS-1998-032

\bibitem{noi}
  V.~P.~Nair, {\tt hep-th/9809086}\\
  O.~Philipsen, {\tt hep-ph/9809436}


 \end{thebibliography}
\end{document}